\documentclass[sigconf,nonacm]{acmart}
\AtBeginDocument{%
  }

\copyrightyear{2025}
\acmYear{2025}
\acmConference[PACT]{Parallel Architectures and Compilation Techniques'25}{November 3-6, 2025}{Irvine, CA}
\acmISBN{978-1-4503-XXXX-X/2018/06}
\setcopyright{none}
\usepackage{amsmath,amsfonts}
\usepackage{algorithmic}
\usepackage{graphicx}
\usepackage{textcomp}
\usepackage{xcolor}
\usepackage{multirow}
\usepackage{xurl}
\usepackage{array}
\usepackage{pifont}
\begin{document}

\title{SCREME: A \underline{Sc}alable Framework for \underline{Re}silient \underline{Me}mory Design}

\author{Fan Li}
\affiliation{%
  \institution{College of Engineering and Computer Science, University of Central Florida}
  \country{USA}}
\email{fan.li@ucf.edu}

\author{Mimi Xie}
\affiliation{%
  \institution{College of AI, Cyber and Computing, University of Texas at San Antonio}
  \country{USA}}
\email{mimi.xie@utsa.edu}

\author{Yanan Guo}
\affiliation{%
  \institution{Department of Computer Science, University of Rochester}
  \country{USA}}
\email{yanan.guo@rochester.edu}

\author{Huize Li}
\affiliation{%
  \institution{College of Engineering and Computer Science, University of Central Florida}
  \country{USA}}
\email{huize.li@ucf.edu}

\author{Xin Xin}
\authornote{Corresponding author. 
Accepted at PACT 2025. 
This is the author's version of the work. 
The definitive Version of Record will appear in 
\textit{Proceedings of the 34th ACM/IEEE International Conference on Parallel Architectures
and Compilation Techniques.}}
\affiliation{%
  \department{College of Engineering and Computer Science}
  \institution{University of Central Florida}
  \country{USA}}
\email{xin.xin@ucf.edu}

\begin{abstract}

The continuing advancement of memory technology has not only fueled a surge in performance, but also substantially exacerbate reliability challenges. Traditional solutions have primarily focused on improving the efficiency of protection schemes, i.e., Error Correction Codes (ECC), under the assumption that allocating additional memory space for parity data is always expensive and therefore not a scalable solution.

We break the stereotype by proposing an orthogonal approach that provides additional, cost-effective memory space for resilient memory design. In particular, we recognize that ECC chips (used for parity storage) do not necessarily require the same performance level as regular data chips. This offers two-fold benefits: First, the bandwidth originally provisioned for a regular-performance ECC chip can instead be used to accommodate multiple low-performance chips. Second, the cost of ECC chips can be effectively reduced, as lower performance often correlates with lower expense.
In addition, we observe that server-class memory chips are often provisioned with ample, yet underutilized I/O resources. This further offers the opportunity to repurpose these resources to enable flexible on-DIMM interconnections.
Based on the above two insights, we finally propose SCREME, a scalable memory framework leverages cost-effective, albeit slower, chips --- naturally produced during rapid technology evolution --- to meet the growing reliability demands driven by this evolution.


\end{abstract}

\settopmatter{printacmref=false}
\settopmatter{printfolios=false}
\maketitle

\section{Introduction}

High-performance computing (HPC) systems often adopt additional protection schemes, i.e., Error Correction Code (ECC), for memory to ensure data correctness. However, the existing architectural support for ECC is becoming insufficient to meet the ever-growing ECC requirement due to the lack of a flexible interface that allows scalable and adaptable ECC configurations.

The necessity for scalable and adaptable ECC configurations stems from two perspectives. First, with the advancement of memory technology, memory subsystems are facing growing reliability challenges due to increasingly complicated manufacturing processes and rising data transmission rates~\cite{hong2010memory, hassan2023improving}. Traditional strategies, such as imposing conservative timing margins to accommodate a few weak cells, have become inadequate for today’s DRAM manufacturing demands. In this context, ECC is beyond the scope of merely preventing unpredictable errors and becomes a crucial factor in determining an optimal performance-resilience trade‑off. Consequently, the memory industry is integrating ECC more extensively across the memory hierarchy, including chip-level (on-die), rank-level ECC, and even inter-DIMM-level ECC. As DRAM technology continues to evolve, the role of ECC in ensuring both performance and reliability will only grow more important.

Second, memory reliability challenges are further exacerbated in HPC environments. Driven by surging memory-intensive AI applications, the capacity of higher-performance memory subsystems is expanding rapidly, where even rare DRAM faults can have an outsized impact on overall system reliability. For example, large data centers that commonly contain millions of DRAM chips can experience memory failure within mere hours~\cite{sridharan2012study}. Even if the per‑bit error rate remains the same in future memory technologies, extending the time-to-failure from hours to days with a $5\sim10\times$ increase in capacity would require approximately two orders of magnitude improvement in per-chip reliability.

Due to its significance, we introduce \textit{SCREME}, a scalable framework for resilient memory design that tackles two major obstacles hindering ECC advancement: (i) the rigid, non-extensible interface (i.e., JEDEC DDR standard) and (ii) expensive storage overhead associated with parity data.
This is in stark contrast to previous studies that typically focused on boosting the efficiency and utilization of existing ECC resources. For example, prior work has shown that exposing on-die ECC resources to the rank-level ECC can significantly improve overall ECC efficiency~\cite{gong2018duo, nair2016xed}, while other studies have explored advanced ECC algorithms to diminish redundancy overhead~\cite{manzhosov2022revisiting, kim2023unity} and enhance error correction capability~\cite{bamboo, zhang2018exploring}. Orthogonal to these efforts, SCREME advocates a feasible protection mode that can allow ECC enhancement on another front, i.e., facilitating efficient utilization when cost-effective parity storage is available.

In particular, we observe that the default practice of allocating identical memory resources to regular and parity data is not always necessary. Parity data can instead be stored in low-performance DRAM chips with reduced channel bandwidth. This strategy is viable because the portion of parity data assigned for error correction is often unnecessary to fetch, given that errors rarely occur. Consequently, with less data needing to be transmitted, ECC chips in commodity modules are indeed over‑provisioned with bandwidth. This, in turn, allows us to accommodate more ECC chips within the same channel width by using slower DRAM chips for parity storage.
Moreover, to enable seamless integration of additional ECC chips, we introduce a configurable architecture that taps into the excess internal I/O bandwidth already provisioned in server-class DRAM chips. In practice, memory designers often allocate more I/O resources than needed to reduce design complexity, unleashing the opportunity to repurpose these excess resources for flexible chip interconnections. This further augments the integrated ECC chips with a range of functions under different I/O configurations. For example, an ECC chip can be held in reserve (as a spare chip) to replace a failed chip on the fly.
Based on the above designs, SCREME can accommodate multiple ECC chips on a single memory module without altering the existing interface.


Beyond technical innovations, we emphasize that memory advancements should also be grounded in market realities. To cut parity storage costs, SCREME explores the use of slower, low-performance chips as a more affordable option. In particular, based on DDR market data~\cite{statista_dimm_market_2018, carnevale_ddr5_2021, fisher1971simple, grubler1990rise}, we observe that with the introduction of higher-speed DDR chips, earlier slower chips often experience a sharp price decline. Furthermore, these slower chips maintain good availability, as they continue to represent a significant portion of shipments due to substantial inventories persisting within the lengthy global supply chain (see Section~\ref{sec: economy}). This makes slow chips both practical and economical for parity storage.

For the rest of the paper, we briefly discuss the background and motivation in Section 2. We elaborate our primary design, which includes three subdesigns in Section 3. We present the evaluation results and conclusion in Section 4 and Section 5, respectively.

\section{Background}

\subsection{Commodity Memory Organization}
DRAM memory is structured in a hierarchical architecture, from DIMMs (Dual In-Line Memory Modules) down to in-DRAM banks\cite{chatterjee2017architecting, yoon2011adaptive, yoon2012dynamic, o2017fine}. 
Each DIMM consists of multiple ranks, and each rank is composed of several DRAM chips. 
A key advancement in DDR5 DIMMs over earlier generations is the inclusion of two independent channels per DIMM, whereas previous generations featured only one.
To achieve larger memory capacities, high-performance memory often uses DRAM chips with narrower I/O width to meet the channel's bit width requirement.
In data centers, for example, a typical channel configuration uses 10 $\times$4 chips, consisting of 8 data chips and 2 parity chips (Figure~\ref{fig: background}(d)). 
Each chip has 4 data pins, allowing 32 bits from the data chips to be transferred in one operation, known as a beat.
Transferring a typical data block, i.e., one cacheline (64B), requires 16 beats (Figure~\ref{fig: background}(e)){\cite{kim2015bamboo}.

\label{sec: background}


\begin{figure}[htbp]
  \centering
  \vspace{-1em}
  \includegraphics[width = 0.45 \textwidth]{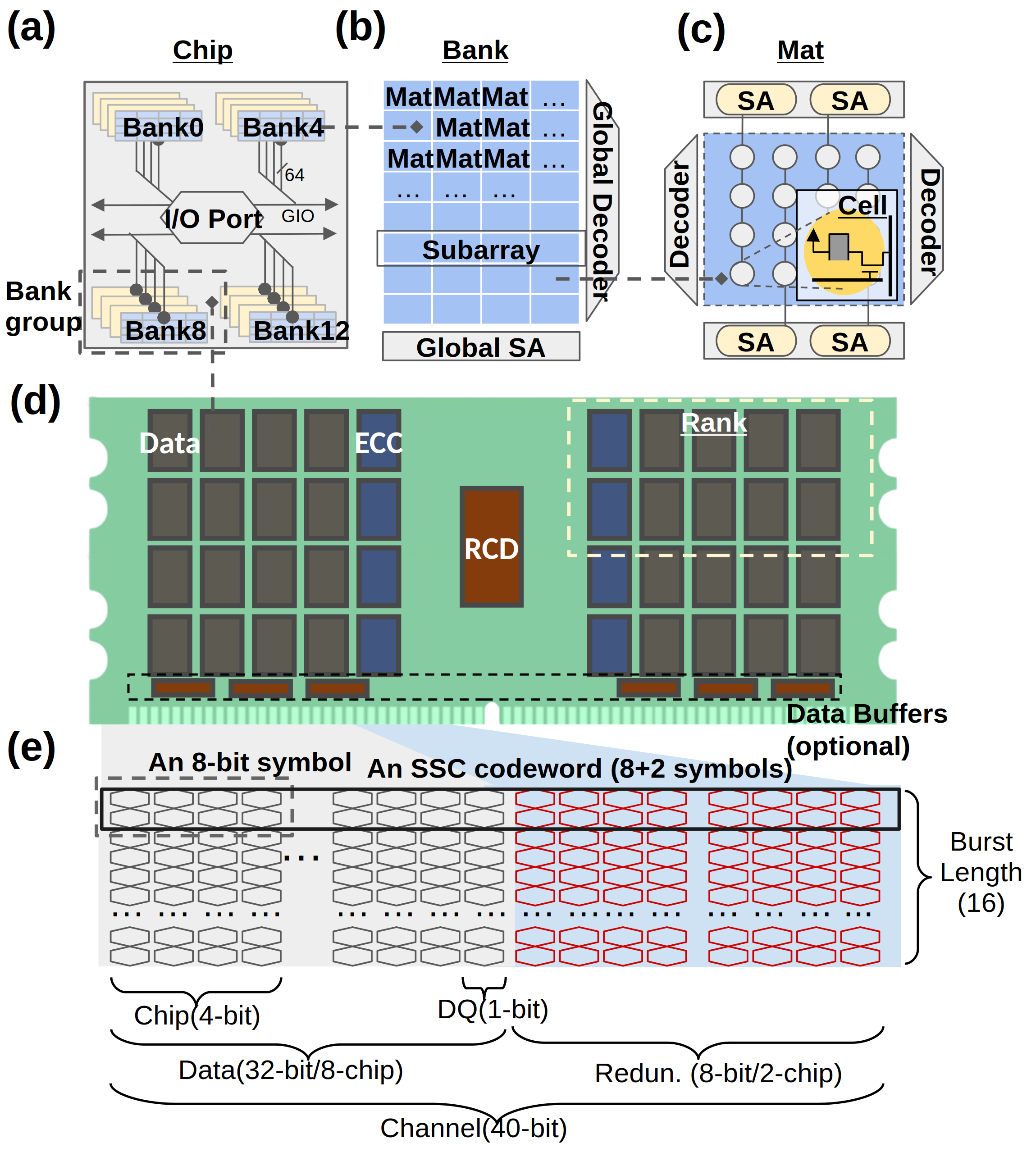}
          \vspace{-1em}
  \caption{\textbf{(a) - (d)} Chip, Bank, Mat, and DIMM.
  \textbf{(e)} A memory transfer block with ChipKill ECC.
  }
  \label{fig: background}
        \vspace{-0.5em}
\end{figure}

Each DRAM chip consists of multiple banks. 
In recent DRAM generations~\cite{DDR4_1, DDR4_2, DDR4_3, jedec-ddr4, jedec-ddr5}, the banks are grouped into {\em bank groups} with each bank group typically having four banks and an independent GIO bus connecting to the I/O port (Figure~\ref{fig: background}(a)). 
Bank groups have semi-independent control logic, allowing for shorter timing intervals between consecutive operations from different bank groups compared to those from the same group.
Within each bank, there are multiple subarrays (Figure~\ref{fig: background}(b)) connected by global bitlines (BLs) and column select lines (CSLs).
Each subarray consists of multiple mats, and each mat forms a 2D cell array (Figure~\ref{fig: background}(c)). 
Inside a mat, cells are vertically connected by local BLs to local sensing amplifiers (SAs), also known as row buffers, and are horizontally connected by local wordlines (WLs)~\cite{Zhang_ISCA2014, micron}.

\subsection{Memory Protection Schemes -- ChipKill} 
\label{sec:chipkill}

DRAM memory achieves high density at the cost of reliability, necessitating additional protection schemes, i.e., Error Correction Code (ECC), to ensure data correctness. Memory ECC schemes are structured hierarchically to align with the organization of memory itself, including Rank-level and Chip-level (on-die) ECC. High-performance memory typically employs advanced Chip-kill ECC at the Rank-level and simpler Single-bit Error Correction (SEC) schemes for on-die ECC.

Chip-kill ECC is typically based on the Reed-Solomon (RS) code, which operates on a set of bits, termed a symbol, instead of individual bits~\cite{reed1960polynomial}. The single symbol error correction (SSC) is often the default RS code for ChipKill ECC~\cite{AMDchipkill, HPchipkill, kim2015bamboo, ECC1, ECC2}. As illustrated in Figure~\ref{fig: background}(e), this scheme treats eight data bits (spanning two beats) from each chip as a symbol. 
By using two check symbols from two ECC chips (a.k.a., parity chips), SSC can correct a single symbol error. This entails that it can recover from the failure of one defective DRAM chip.

The significant challenge of ChipKill lies in its increasing overhead. Since the SSC algorithm always requires two check symbols (i.e., two ECC chips), a decrease in channel width leads to a higher proportion of ECC chips. In particular, traditional DDR4 utilizes a 72-bit channel comprising 16 data chips and 2 ECC chips, presenting 12.5\% redundancy overhead. In DDR5, with a 40-bit channel width, the redundancy doubles to 25\% as it includes 8 data chips and 2 ECC chips. More significantly, DDR6 is anticipated to use a 24-bit channel width \cite{jedec_mobile_2024}. This induces a 50\% overhead for ChipKill implementation, which is considered `unaffordable'.



\subsection{I/O Configurations -- Common Die}
\label{sec:IO}


As shown in Figure~\ref{fig: IO}, modern memory chips provide I/O widths of 4 ($\times$4), 8 ($\times$8), or 16 ($\times$16) bits for different deployment scenarios{~\cite{synopsys}}, so that each chip transmits 4, 8, and 16 bits per beat, referred to as $\times$4, $\times$8, $\times$16 configurations, respectively. Typically, a narrower I/O width supports ECC schemes with higher efficiency. As a result, different levels of ECC protection correspond to different I/O width configurations. For instance, $\times$16 chips are generally used without ECC, $\times$8 chips are commonly paired with SEC-DED, and $\times$4 chips are employed with stronger Chip-Kill ECC{\cite{dell1997white, schroeder2009dram}}. In this paper, we focus on server-class memory that often employs $\times$4 chips to meet the high-performance and reliability requirements.

To mitigate the cost of design, DRAM manufacturers adopt the \textbf{common die} design for different I/O configurations{~\cite{patent1, patent2, patent3, patent4, ERUCA, xin2021sam, cir1, samsung}}, that is, the maximum of 256-bit internal bus and the associated 256-bit I/O buffer (16-bit for each of the 16 pins) are integrated in all DRAM chips. The desired configuration is implemented by cutting electric fuses after passing tests for a specific I/O mode. For example, a $\times$4 configuration activates one-fourth of the internal bus and I/O buffer resources to store 64 bits of data fetched from the DRAM arrays (Figure~\ref{fig: IO}(a)). In contrast, a $\times$8 configuration activates half of these internal resources to store 128 bits of fetched data (Figure~\ref{fig: IO}(b)). Interestingly, $\times$4 and $\times$8 configurations even share a common chip package, as shown in Figure~\ref{fig: IO}(d).
The common die design not only mitigates the design, verification, and tooling costs, but also adapts better to market uncertainty. Of course, it leaves a subset of I/O buffers, as well as the associated internal bandwidth, unused under $\times$4 and $\times$8 configurations.

\begin{figure}[htbp]
  \centering
  \vspace{-0.15in}
  \includegraphics[scale=0.45]{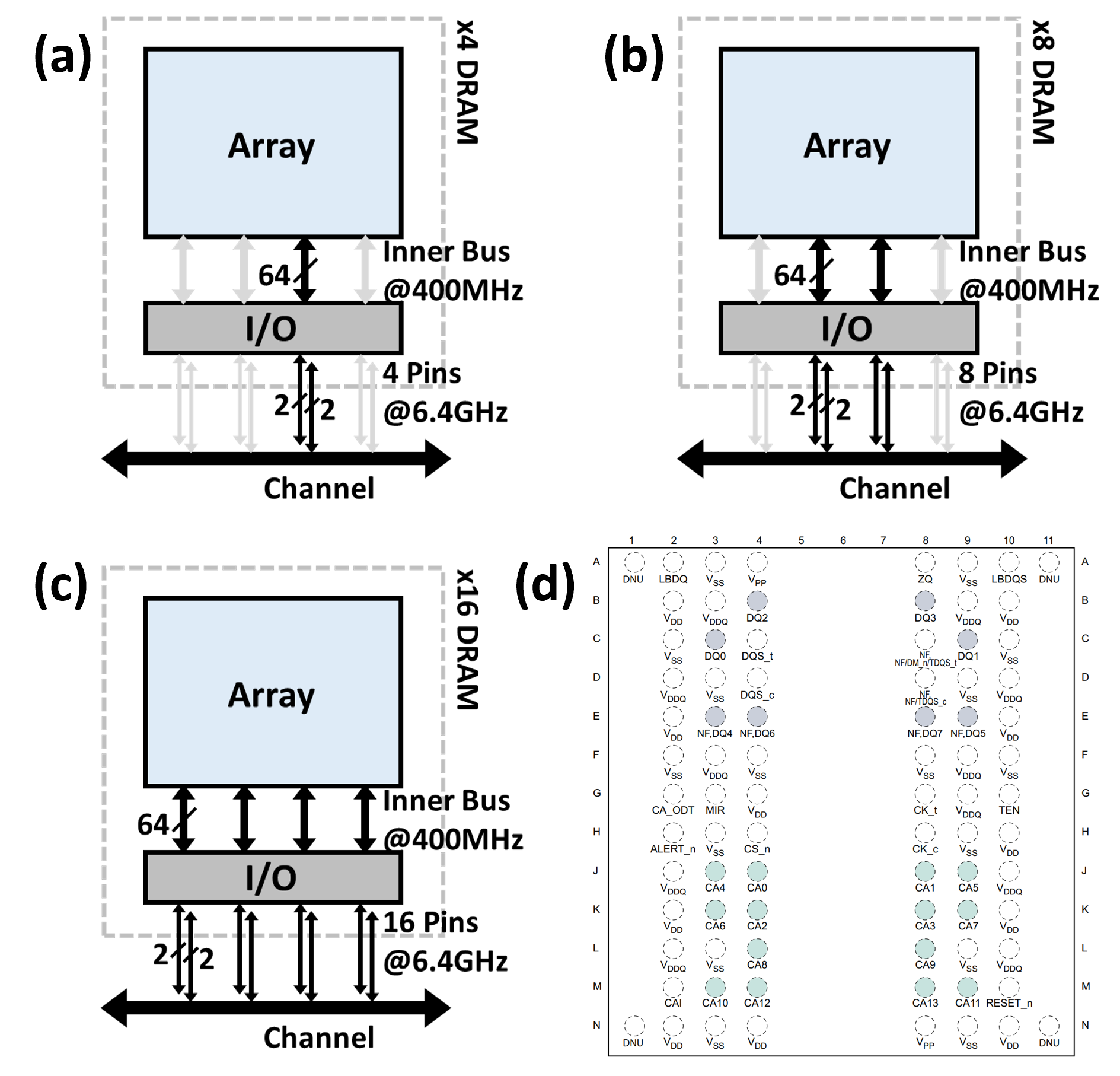}
  \vspace{-0.15in}
  \caption{\textbf{(a) - (c)} DRAM chips with I/O widths of 4 ($\times$4), 8 ($\times$8), and 16 ($\times$16). 
  \textbf{(d)} pin layout for both $\times$4 and $\times$8 configurations
  }
  \label{fig: IO}
\end{figure}


\subsection{Slow‑Chip Economy}
\label{sec: economy}

DRAM technology has seen continuing advancement, characterized by the rapid growth of transmission rates. As shown in Figure~\ref{fig: market}(a), the adoption of higher-speed memory follows a steep trajectory\footnote{modeled using the Fisher-Pry substitution model based on historical DDR bit shipment data \cite{fisher1971simple, carnevale_ddr5_2021}.}, demonstrating a rapid replacement of preceding modules within relatively short time frames.
Accordingly, memory chip prices often experience steep declines with the emergence of successive products. For instance, as shown in Figure~\ref{fig: market}(b)\cite{carnevale_ddr5_2021}, a predecessor memory chip typically costs only half as much as its successor, whereas the speed reduction (i.e., transmission rate) is only around 800$\sim$1600 MT/s. The cost difference becomes even more pronounced with greater speed reduction. Note that the terms "predecessor" and "successor" do not refer to different DDR generations. Instead, they represent product speed advancements either within the same generation or across different generations.

On the other hand, the rapid development of advanced chips does not necessarily lead to the disappearance of predecessor chips. As shown in Figure~\ref{fig: market}(a), shipments of lower-performance, predecessor chips remain significant when the speed reduction is relatively small. This is largely due to the inherent complexity and inertia of global semiconductor supply chains, which constrain their ability to quickly adapt to abrupt shifts in demand. Intuitively, the smaller the performance gap between legacy chips and newer models, the more significant the inventories that remain within the supply chain.
Our proposal is to leverage the available, low cost predecessor chips to enhance memory module resilience. This is important as we are still in the early stage of the DDR5 era, where the rated frequencies (currently around 4800$\sim$5600 MT/s) are expected to grow significantly  in the near future (potentially exceeding 10,000 MT/s)\cite{rambus2024chipsets}. Preparing current DDR5 chips for potential integration into future DDR5 modules is critical for hardware efficiency and sustainability.\footnote{In contrast, our proposal does not involve recycling already outdated chips into current modules, as such chips cannot be augmented with our designs and often have incompatible interfaces.} This strategy may also facilitate cross-generation adoption, allowing DDR5 chips to be recycled within DDR6 systems.

\begin{figure}[htbp]
  \centering
  \vspace{-0.15in}
  \includegraphics[width=0.5\textwidth]{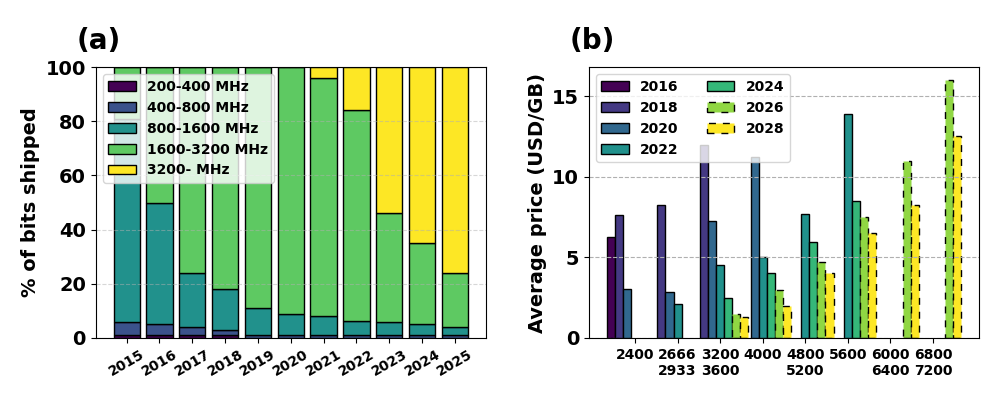}
  \vspace{-0.3in}
  \caption{Shipment share (a) and average unit price (b) of DDR across different data rates and years. Dotted bars indicate projected values.}
  \label{fig: market}
\end{figure}

\subsection{Various Failure Challenges}
\label{sec: challenges}

Many prior studies on memory reliability have focused on small-scale internal faults. However, recent work~\cite{sridharan2012study, beigi2023systematic, schroeder2009dram, sridharan2015memory, cha2017defect} has shown that device-level failures are not uncommon. Moreover, to meet stringent reliability requirements, server-class memory chips are often retired before they exhibit complete failure. Figure~\ref{fig: prob} summarizes three types of device-level failures, i.e., single-chip, data-wire, and control-wire failures, each of which can impact all associated ranks within a module.
Taking single-chip failure as an example, it leads to two major issues. First, the effectiveness of ECC is compromised. For example, in a rank where one chip has already failed, ChipKill ECC can no longer guarantee chip-level protection if another chip encounters transient multi-bit faults. Second, a failing chip can reduce the usability of otherwise functional chips within the same rank. To avoid the waste of the entire module in such cases, it requires sophisticated management strategies to isolate or circumvent the failure.

To address these device-level failure challenges, traditional approaches opt to use advanced protection schemes that encompass a larger number of chips within a single ECC codeword. For example, Intel introduced Double Device Data Correction (DDDC)~\cite{supermicro2017memoryras}, which protects against two \textit{sequential} chip failures. DDDC requires a combination of two channels in a lockstep manner, such that four ECC chips (with two ECC chips per channel) are involved in each memory transfer. DDDC uses three of these ECC chips to implement protection, and the remaining one is used as a spare chip. The controller continuously counts corrected errors for each chip. If the count of one chip reaches a certain threshold, this chip is regarded as a failing chip and is replaced by the spare chip. 
However, the lockstep operation of two channels often introduces substantial performance overhead, which we analyze in the evaluation section.

\begin{figure}[htbp]
  \centering
  \vspace{-0.15in}
  \includegraphics[width=0.23\textwidth]{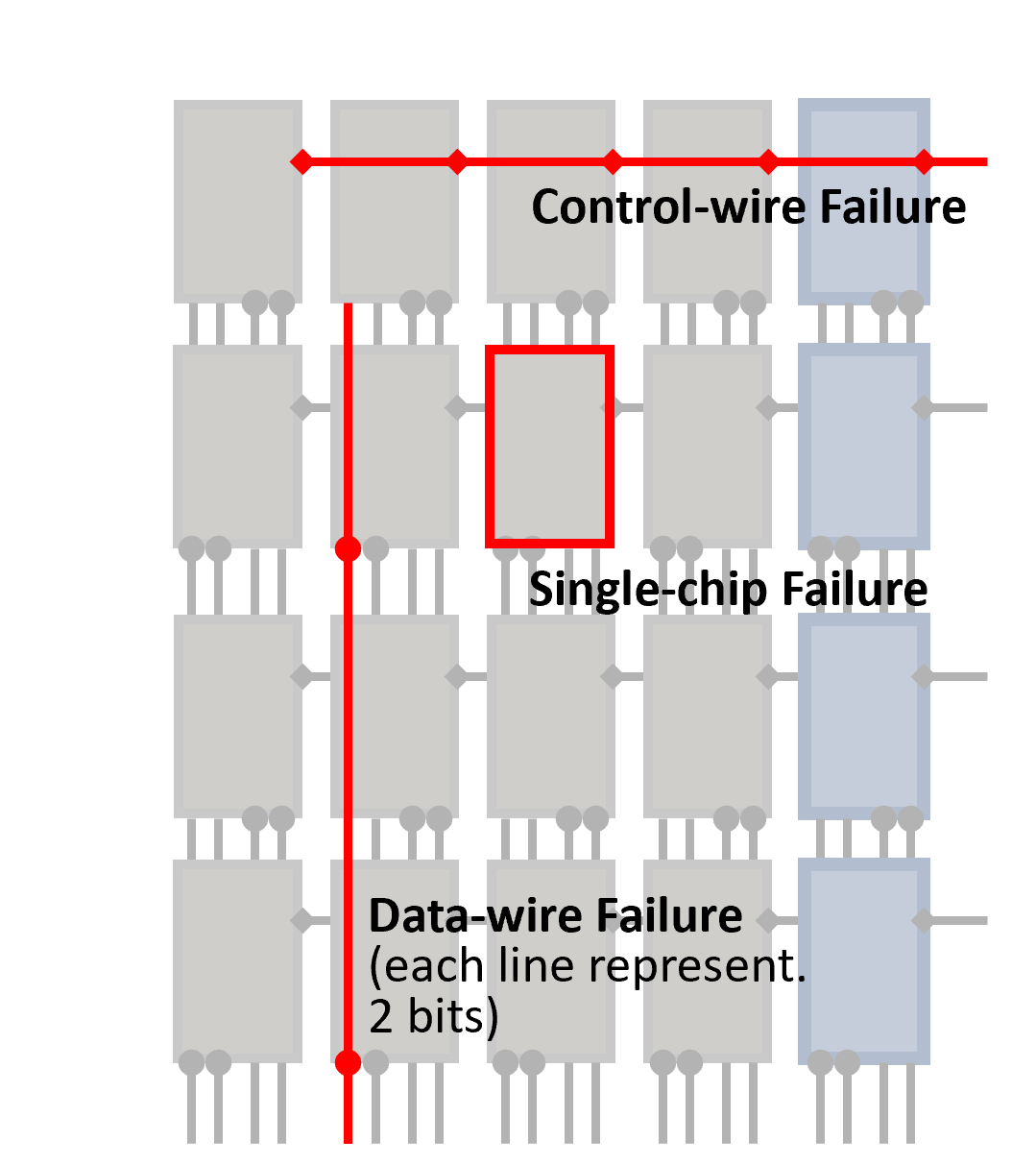}
  \vspace{-0.05in}
  \caption{Three types of device-level failures.
  }
  \vspace{-0.15in}
  \label{fig: prob}
\end{figure}





\section{SCREME Design}
\label{sec: design}

We envision a forward‑looking framework for resilient memory design that transcends traditional focuses on maximizing ECC efficiency and instead extends extra room for broader ECC deployment. In this section, we first introduce how to utilize a slow chip as an ECC chip in a modern memory module. Next, we present a configurable I/O design that enhances the module's resilience to device-level failures. Finally, we propose a comprehensive framework that integrates these two designs.


\subsection{Write-Only ECC Chips}
\label{sec: write-only}

Unlike conventional memory modules composed solely of identical chips, SCREME incorporates slower chips into modern modules with negligible performance overhead. This hybrid module is founded on the insight below:

\textbf{Insight 1: ECC chips do not require the same level of performance as data chips.} Specifically, the portion of parity data assigned to error correction is actually not needed until an error is detected. Given that errors rarely occur, this parity data for error correction can be retained in ECC chips without transmission during read accesses. For the Chipkill ECC (see Section~\ref{sec:chipkill}) that incorporates two ECC chips, we can employ only one ECC chip to achieve the same error detection capability. The bandwidth that is allocated to the remaining ECC chip can be repurposed during read accesses. 

Inspired by the observation, we allocate half of the ECC bandwidth to accommodate low-performance chips. As shown in Figure~\ref{fig: write-only}(a), the last ECC chip is replaced with a slow chip, which is characterized by its lower transmission rate. This chip is dedicated solely to write accesses. Its extended write time resulting from the reduced speed can be overlapped when other chips in the same rank perform read accesses (see Section~\ref{sec:ECC flow}). Given the relatively small portion of write accesses\footnotemark, 
\footnotetext{A store instruction that misses the cache first triggers a memory read access, thereby theoretically limiting memory write accesses to at most 50\%.}
the write overhead of this slow chip can be mostly amortized, resulting in a negligible impact on overall performance. 
It is worth noting the following two points. First, ``slow chips” in the paper refers to newly manufactured, lower‐speed chips available in the market, instead of used ones. Second, operating a memory chip below its rated frequency is always safe and even provides additional timing margins~\cite{chang2017understanding}. Therefore, a low-performance chip (e.g., rated at 4800 MT/s) can be conservatively operated at 3200 MT/s. Consequently, a memory channel running at 6400 MT/s can easily synchronize with the slower chip by simply transmitting each bit twice, eliminating the need for clock adjustments on the memory controller side.

\begin{figure}[htbp]
  \centering
  \vspace{-0.15in}
  \includegraphics[width=0.38\textwidth]{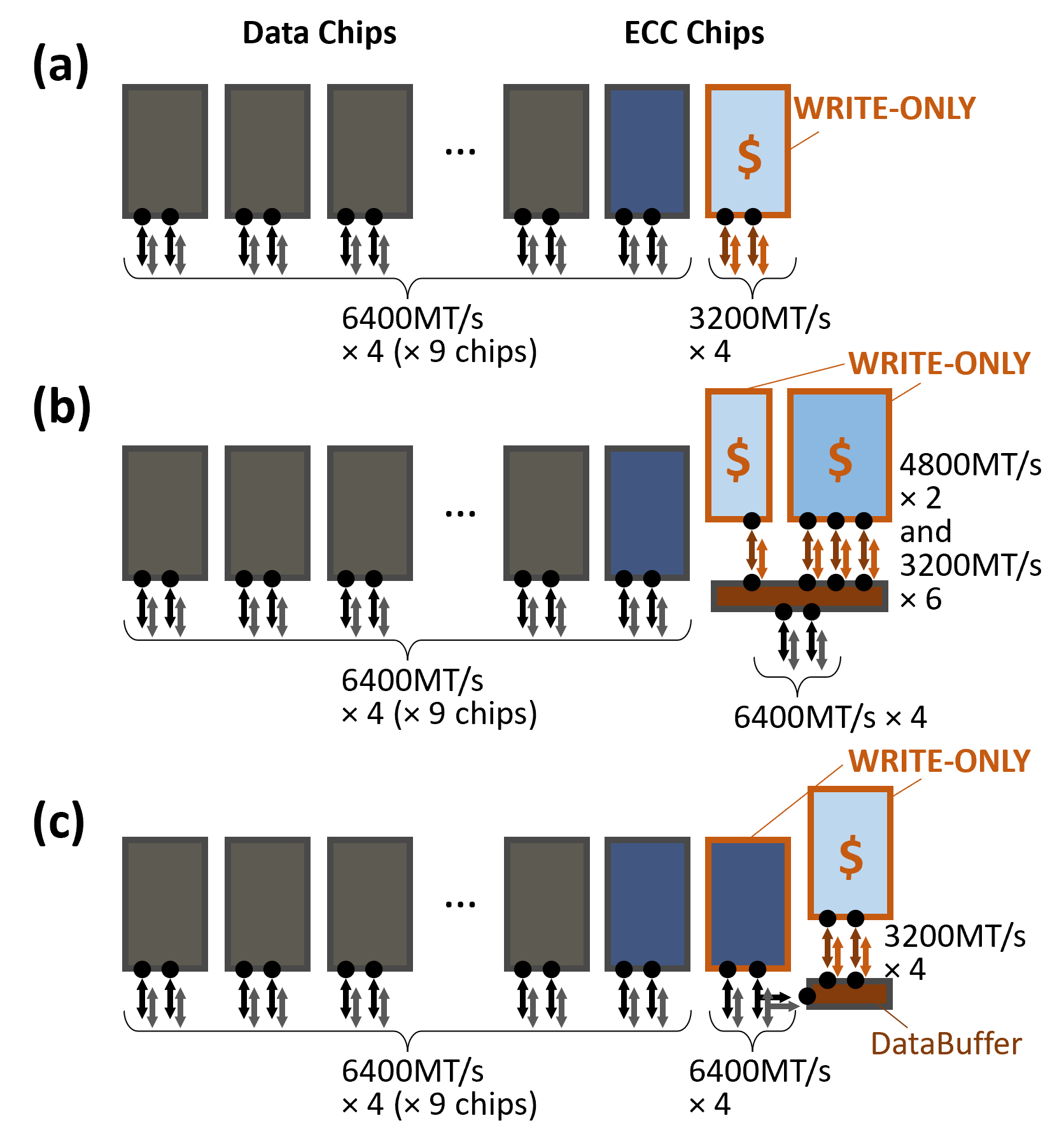}
  \vspace{-0.15in}
  \caption{Design of SCREME-WO: \textbf{(a)} Employ a slow chip as an ECC chip.
  \textbf{(b)} Employ various slow chips to store parity data.
  \textbf{(c)} Employ a slow chip as a spare ECC chip.
  }
  \label{fig: write-only}
\end{figure}

Since low-performance chips typically come at a lower price point (see Section~\ref{sec: economy}), the write-only configuration in Figure~\ref{fig: write-only}(a), termed \textit{SCREME-WO}, can reduce overall module cost by replacing one expensive ECC chip. To further enhance the benefits of this configuration, we introduce two implementation strategies. First, we employ a data buffer, a widely used component in LRDIMM~\cite{mouser_ddr5_rdimm} and MCR-DIMM~\cite{skhynix_mcr_dimm, micron_mrdimm_brief}. It buffers data from the channel and modulates the data transmission rate to match that of slow chips~\cite{zheng2008mini, asghari2016chameleon, yoon2011adaptive}. Additionally, it supports integration of multiple chips by extending the number of I/O lines, enabling flexible capacity and performance configurations. For instance, as illustrated in Figure~\ref{fig: write-only}(b), two chips with different capacities can be used jointly by assigning proportional I/O widths. While the data buffer might induce extra access latency, it does not impact overall performance since this extra write latency applies only to write operations, which are not on the critical path of program execution.
Second, in addition to replacing one ECC chip, the write-only configuration provides a low-cost method to involve spare chips in a memory module. As shown in Figure~\ref{fig: write-only}(c), the spare chip is only activated when needed, such as providing additional protection when encountering increased reliability challenges. This spare chip shares a 2-bit channel width with the last ECC chip through a data buffer. When activated, both the spare chip itself and the last ECC chip are configured as write-only. 

It is also worth noting that our research emphasizes forward-looking resilient memory architectures rather than recycling legacy chips into current modules. For example, recycling DDR4 chips into DDR5 modules is not the focus of SCREME, as their incompatible interfaces can diminish any potential benefit. Instead, during the early stages of the DDR5 era, it is crucial to ensure that current memory chips (e.g., DDR-4800 or DDR-5600) are well-prepared to support future high-speed modules (e.g., DDR-8000).
\textcolor{black}{More importantly, our research prepares a sustainable solution for next-generation DDR6 modules where the channel width is further reduced from 40 bits (in DDR5) to 24 bits (see Section~\ref{sec:chipkill}). In this setting, using two ECC chips for ChipKill implementation introduces a 50\% overhead relative to the four data chips. A cost-effective solution becomes increasingly critical for ECC scaling.}

However, the designs in Figure~\ref{fig: write-only} provide only error detection capabilities, necessitating additional handling for exceptional cases when errors are identified. We defer the detailed discussion to Section~\ref{sec:ECC flow}. Furthermore, some of these designs require flexible I/O configurations. This motivates the I/O design in the following subsection.

\begin{figure}[htbp]
  \centering
  \vspace{-0.15in}
  \includegraphics[width=0.43\textwidth]{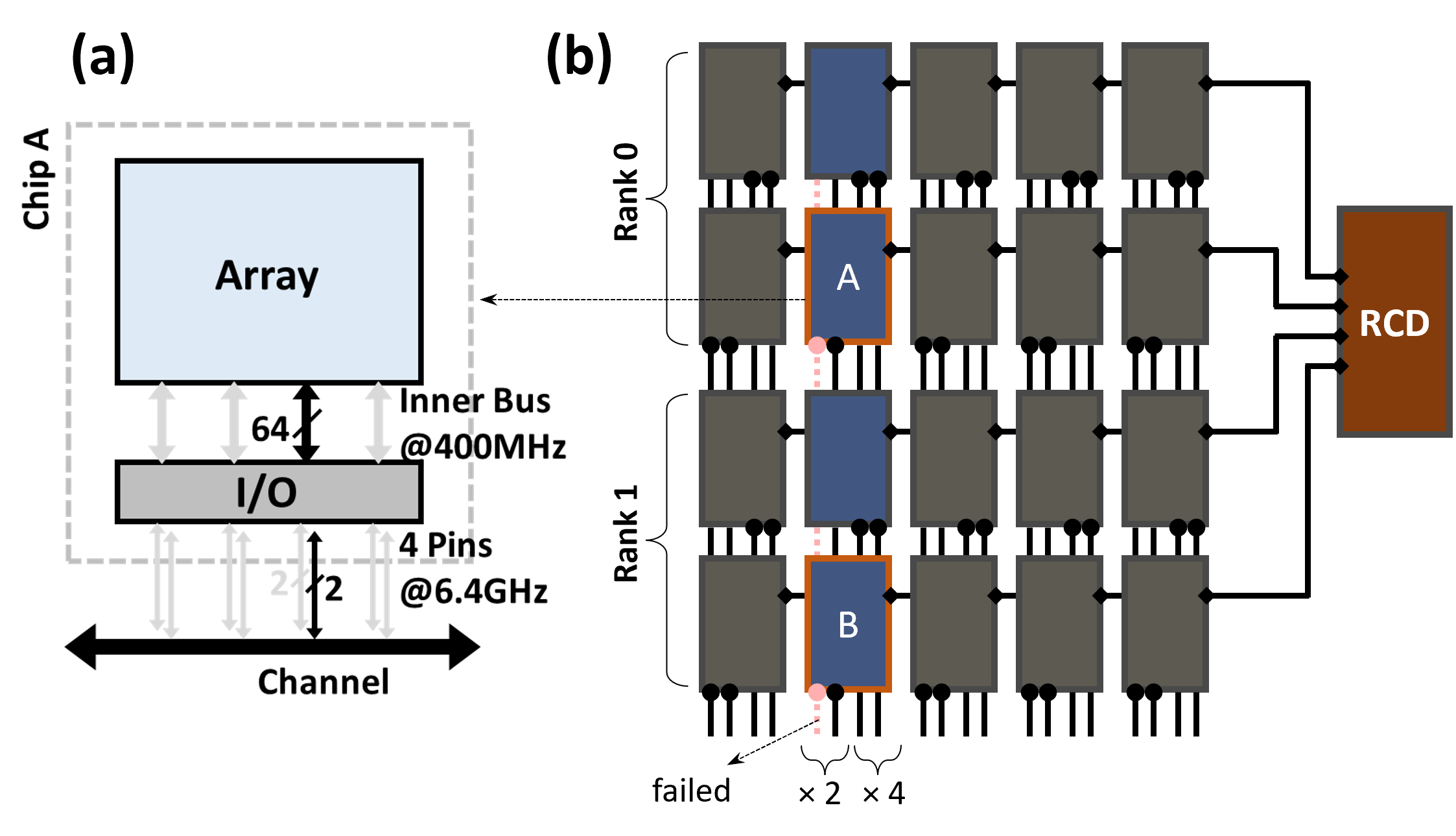}
  \vspace{-0.1in}
  \caption{Design of SCREME-I/O(col): \textbf{(a)} I/O configuration details.
  \textbf{(b)} Tolerate data wire failures.
  }
  \vspace{-0.2in}  
  \label{fig: resil_1}
\end{figure}

\subsection{Configurable I/O Interface}
\label{sec: life}

Traditionally, DRAM chips in a memory module have a fixed, unchangeable I/O width. This inflexibility not only restricts the use of more cost-effective chip options but also hampers its resilience to diverse failure scenarios, as discussed in Section~\ref{sec: challenges}. We observe that flexible I/O configuration is indeed viable by utilizing the latent resources in commodity DRAM:

\textbf{Insight 2: Ample I/O resources in DRAM can provide flexible chip connections.} As aforementioned, DRAM chips with different I/O widths, e.g., $\times$4, $\times$8, or $\times$16, are manufactured via a common die design. The die is prepared with ample resources and is configured to a specific I/O width by deactivating certain portions of these resources during the packaging process. In line with this trend, $\times$4 chips that are commonly deployed in servers can be further reconfigured to a $\times$2 width by disabling additional resources or expanded to a $\times$8 width by utilizing the unused resources.

We advocate for providing end users with access to this configuration rather than having these I/Os permanently enabled or disabled during chip packaging. Meanwhile, commodity DRAM has provided such logic, i.e., the IO gating unit, to direct internal data buses to the enabled I/Os~\cite{xin2021sam, ERUCA}. 
In this subsection, we leverage such configuration flexibility to address the data wire or control wire failures in high-performance server-class memory modules, which typically comprise two or more ranks. Note that this subsection does not involve the use of slow chips.


As shown in Figure~\ref{fig: resil_1}(b), two data wires are permanently failed (each vertical line in the figure represents 2 bits). This impacts all the column-wise chips (Chip A and B) connected to the line. Simply disabling Chip A and B results in a waste of their storage space. More significantly, this compromises the Chipkill correction capability in each rank. 
To tackle this problem, we propose a column-wise failure-resilient design (\textit{SCREME-I/O (col)}) by combining the above two (ECC and I/O) insights. First, if the affected chips are data chips, the associated column of chips will switch roles with the last column of chips, i.e., the ECC chips. Second, the affected chips will be configured to a narrower $\times$2 I/O width (Figure~\ref{fig: resil_1}(a)) to circumvent the failing wires. Meanwhile, these chips will operate as write-only ECC chips to offset the bandwidth loss. Figure~\ref{fig: resil_1} indicates an example where the second-column chips are switched to function as ECC chips. Chip A and B use the unaffected $\times$2 I/O to update parity data only. This configuration allows for up to two data wire failures.

\begin{figure}[htbp]
  \centering
  \vspace{-0.1in}
  \includegraphics[width=0.43\textwidth]{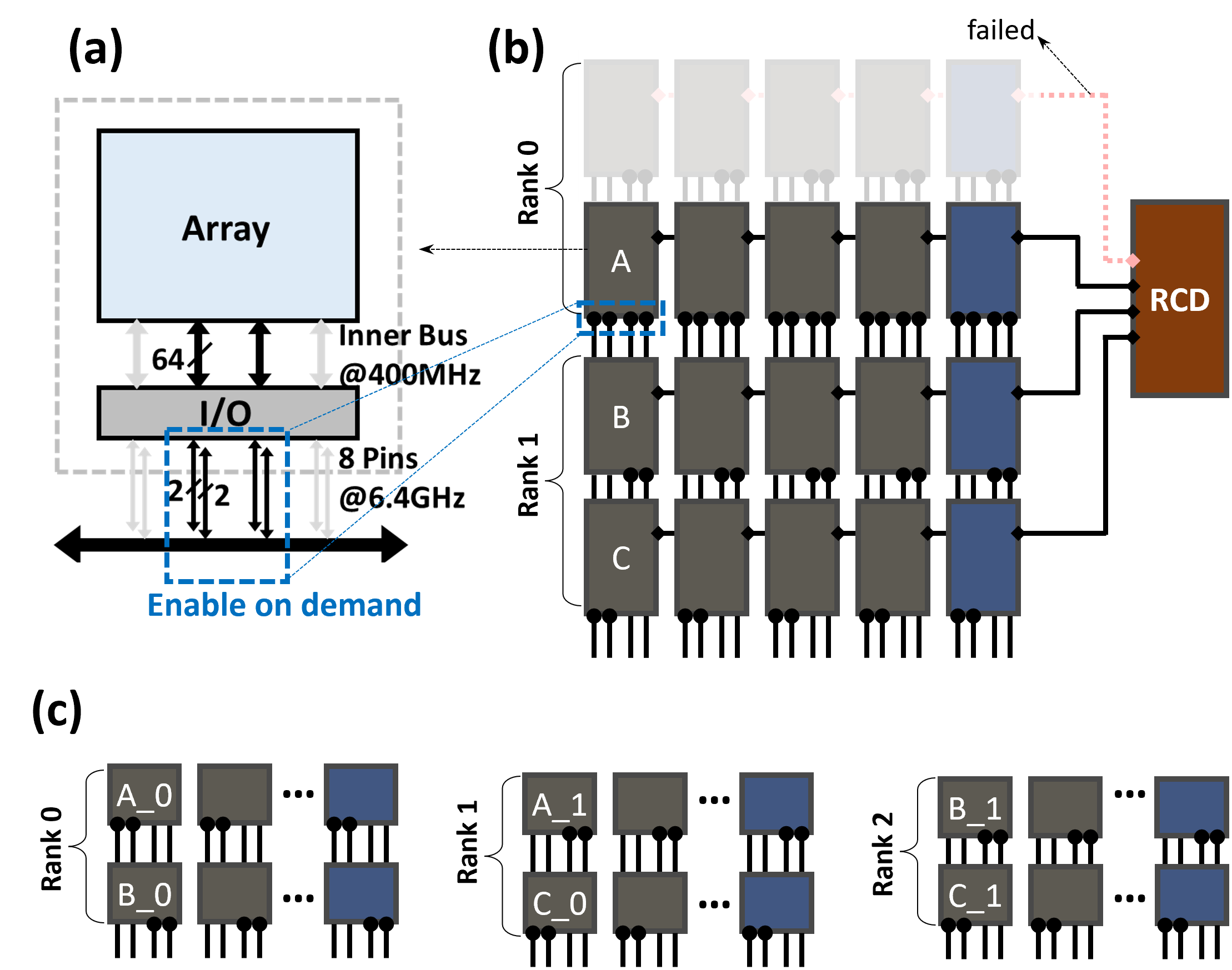}
  \vspace{-0.1in}
  \caption{Design of SCREME-I/O(row): \textbf{(a)} I/O configuration details.
  \textbf{(b)} Tolerate control wire failures.
  \textbf{(c)} Rank reorganization and associated I/O configuration.
  }
  \vspace{-0.1in}  
  \label{fig: resil_2}
\end{figure}

\begin{figure*}[htbp]
  \centering
  \includegraphics[width=0.93\textwidth]{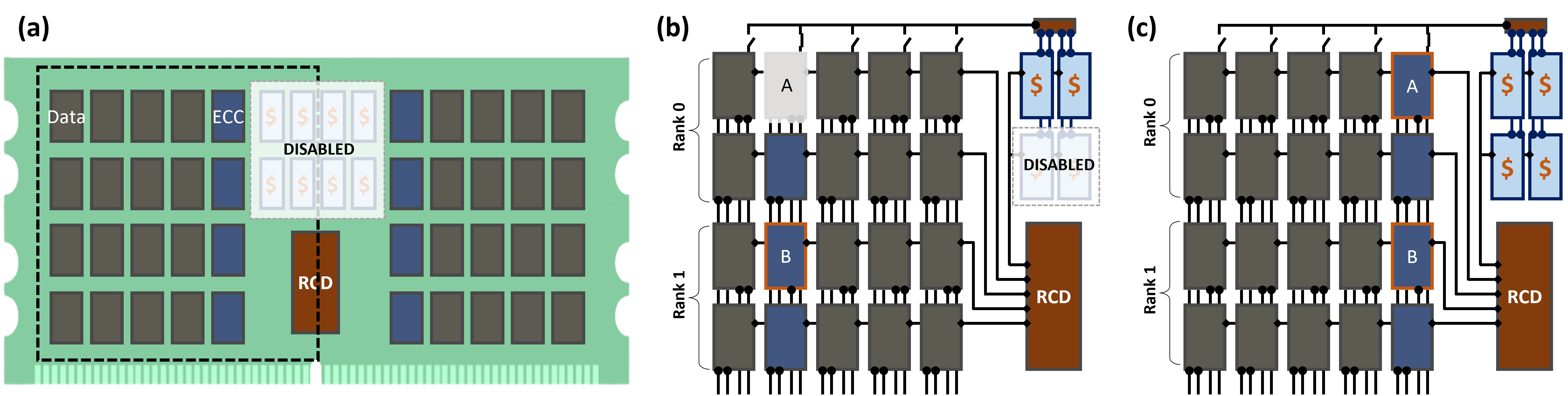}
  \vspace{-0.1in}
  \caption{Design of SCREME-Framewk: \textbf{(a)} The memory module with cost-effective spare chips.
  \textbf{(b)} Chip-replacement mode.
  \textbf{(c)} Scalable-ECC mode.
  }
  \vspace{-0.15in}
  \label{fig: framework}
\end{figure*}


Figure~\ref{fig: resil_2}(b) shows the second scenario, where a permanent failure in a control wire results in the capacity loss of all row-wise chips connected to it. More critically, it also affects the ability to effectively access the remaining functional row within the same rank. For example, if this row is used as a standalone rank, it cannot fully utilize the available channel bandwidth.
To address this problem, we propose a row-wise failure-resilient design (\textit{SCREME-I/O (row)}) by leveraging the underutilized I/O resources in $\times$4 chips. 
Traditionally, data wires are connected to every other chip in the same column. Given that $\times$4 and $\times$8 chips share the same die and package, an alternative is to have these data wires connect to all column-wise chips; however, each chip enables only half of its data pins (i.e., $\times$4 I/O width, as shown in Figure~\ref{fig: resil_2}(a)) to form the same interleaved connection. 

Consequently, if the first row (\textit{Row1}) fails, \textit{Row2} can coordinate with \textit{Row3} and \textit{Row4} to maintain the access efficiency. In particular, chips in \textit{Row2} will enable their left-side $\times$4 I/Os when operating with \textit{Row3} and their right-side $\times$4 I/Os when operating with \textit{Row4}. This effectively transforms the three rows into three separate ranks (Figure~\ref{fig: resil_2}(c)), each comprising half the original capacity. More accurately, the first rank consists of the first half of banks, \textit{Bank 0-15}, in \textit{Row2} and \textit{Row3}. The second rank consists of \textit{Bank 16-31} in \textit{Row2} and \textit{Bank 0-15} in \textit{Row4}. The remaining rank contains \textit{Bank 16-31} in \textit{Row3} and \textit{Row4}. Note that this reorganization can be applied at other hierarchical levels, such as bank groups and subarrays. 
Additionally, this row-wise failure-resilient approach can also accommodate multi-row failures, though these rarely occur. For instance, if \textit{Row1} and \textit{Row3} both fail, the remaining \textit{Row2} and \textit{Row4} can be reorganized into a single rank using the flexible I/O configuration.

The above two I/O-centric designs primarily target commodity memory with external interconnection failure. If a single chip fails due to internal malfunctions, such as multi-bank failure or power unit failure, we can still apply the SCREME-I/O (row) approach to isolate the associated row and to maintain the functionality of the remaining row within the same rank. Yet, this is not an efficient solution as a single chip failure leads to a loss of an entire row. This limitation motivates the next design below. 


\subsection{A Framework for Resilient Memory}

In this subsection, we propose a resilient memory framework, termed \textit{SCREME-Framewk}, by combining both SCREME-WO and SCREME-I/O. 
Figure~\ref{fig: framework}(a) illustrates the architecture of SCREME-Framewk, which integrates a `pool' of slow chips, employed as spare chips and connected to each pair of data wires (each vertical line in the figure represents 2 bits) via an array of switches. By default, the spare-chip pool remains powered off and is only activated when needed. This framework supports two resilient design goals: (1) replacing a single failed chip, and (2) scalable, adaptive ECC.

In particular, if a chip fails permanently, one or a few spare chips in the pool, depending on the density of each slow chip, are activated to provide an equivalent capacity for data storage. This allows the system to maintain the same ChipKill protection as the baseline.
Meanwhile, all the other chips in the same column switch their roles to function as ECC chips, similar to the approach of SCREME-I/O (col) in Figure~\ref{fig: resil_2}. Figure~\ref{fig: framework}(b) indicates a detailed example. Two spare chips with low density are activated to compensate for the capacity loss from the failing chip (Chip A). The other three chips in the second column function as ECC chips. In addition, the four data wires previously routed to both Chip A and Chip B are now evenly divided between Chip B and the spare chips. Specifically, Chip B is reconfigured to a $\times$2 I/O width and repurposed as a write-only ECC chip, dedicated exclusively to updating parity data. The remaining two data wires are connected to the activated spare chips through the associated switch, also dedicated exclusively to parity updating. 

The framework also offers a cost-effective solution for implementing scalable and adaptive ECC, which is necessary for meeting diverse reliability requirements across a range of deployment scenarios. For example, a memory module operating in harsh environments or under low power conditions may require stronger protection. Additionally, emerging challenges, such as rowhammer attacks, impose extra demands on ECC capabilities. As a module begins to experience higher error rates, an effective solution is to upgrade to a more advanced ECC scheme, such as DSD-SSC (Double Symbol Detection and Single Symbol Correction), to counteract the rising errors. 
To accommodate the additional parity space required for such an upgrade, we reconfigure Chip A to a $\times$2 I/O width and allocate the remaining 2-bit channel bandwidth to the spare chips, as shown in Figure~\ref{fig: framework}(c). In this configuration, both Chip A and the spare chips are dedicated exclusively to updating and storing parity data. 

Beyond its resilience features, the framework can be extended to support additional metadata storage. For instance, in DRAM cache architectures, where DRAM serves as a cache in a hybrid DRAM–NVM memory system, storing tags requires extra space and often results in irregular memory access granularity~\cite{qureshi2012fundamental, young2019tictoc}. To address this challenge, SCREME-Framewk can be adapted by designating one ECC chip as a metadata chip, while leveraging spare chips to compensate for the resulting loss in ECC coverage. This configuration mirrors the two resilient designs discussed earlier, in which a 2-bit channel width is allocated to spare chips for parity data updates.


\subsection{Discussion}




It is widely recognized that detecting an error generally requires fewer parity resources than correcting an error. Literature has leveraged this knowledge to enhance memory performance and facilitate innovative ECC designs~\cite{yoon2010virtualized, nguyen2018nonblocking, saileshwar2018synergy}. SCREME contributes a new approach by configuring a ChipKill-ECC chip as `write-only' so as to accommodate a slow chip. This approach introduces two key challenges to the original operation flow of ChipKill: (1) the parity updating process and (2) the ECC decoding process.

\subsubsection{Overlapping Slow Writes}
Typically, memory performs reads and writes in a burst manner (sequences of continuous read or write operations) to mitigate the turnaround overhead incurred when switching from writes to reads. This allows hiding the latency of slow writes behind regular memory operations.
As shown in Figure~\ref{fig: discuss1}, at time 0, both slow and regular chips begin processing write requests. At $t_1$, regular chips complete their writes and switch to serve reads, while the slow chip continues serving writes due to its reduced speed. At $t_2$, regular chips complete their reads and are prepared to switch to process the next round of writes. If the slow chip completes its writes before $t_2$ ($t_1'<t_2$), the slow write overhead is fully overlapped. Otherwise, regular chips are stalled until slow chips complete writes, resulting in a delay of $t_1'-t_2$. In most cases, $t_1'<<t_2$, the overhead introduced by the slow chip during parity updates is negligible.

\begin{figure}[htbp]
  \centering
  \vspace{-0.15in}
  \includegraphics[width=0.4\textwidth]{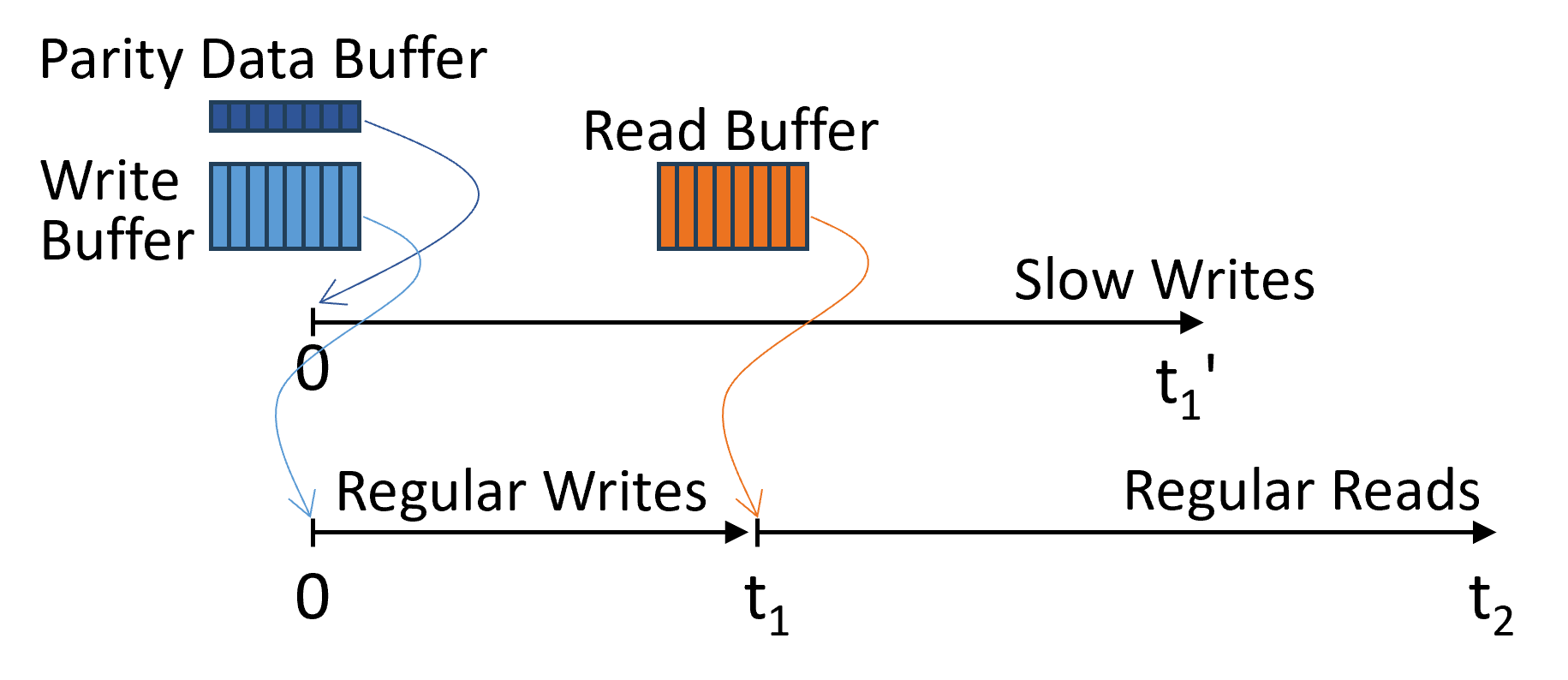}
  \vspace{-0.2in}
  \caption{Runtime overlapping between slow writes and regular writes and reads.
  }
  \vspace{-0.15in}
  \label{fig: discuss1}
\end{figure}

\subsubsection{Operation Flow for ECC}
\label{sec:ECC flow}
Figure~\ref{fig: discuss2}(a) illustrates the default operation flow of ChipKill. When the number of errors exceeds ChipKill’s correction capability, two potential outcomes may arise: (1) the decoder detects the issue and reports a decoding failure  (a.k.a., Detectable but Uncorrectable Errors), or (2) it miscorrects the data, mistakenly assuming successful correction, thereby introducing further errors into the system (a.k.a., Silent Data Corruption). The latter case is particularly problematic, as these undetected errors can lead to severe reliability issues.

\begin{figure}[htbp]
  \centering
  \vspace{-0.10in}
  \includegraphics[width=0.37\textwidth]{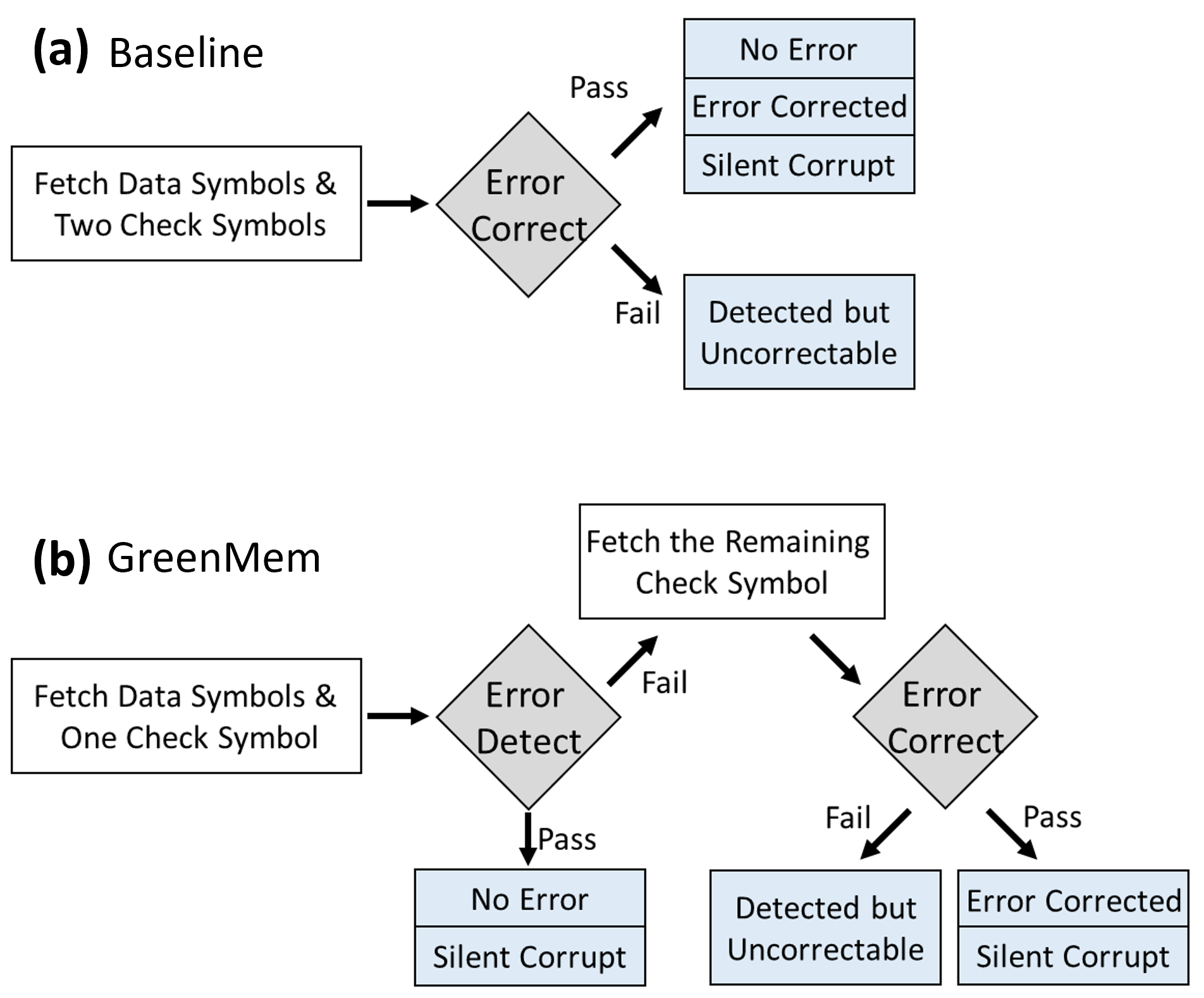}
  \vspace{-0.05in}
  \caption{\textbf{(a)} Operation flow of baseline ChipKill.
  \textbf{(b)} Operation flow of decoupled ChipKill.
  }
  \vspace{-0.15in}
  \label{fig: discuss2}
\end{figure}

SCREME decouples the default error-handling flow into separate detection and correction processes. As shown in Figure~\ref{fig: discuss2}(b), it performs error detection first. If an error is detected, the write-only slow chip is interrupted and temporarily switched to read mode to supply the remaining parity data. During this process, all other chips are stalled to prioritize the transfer of this parity data. Once all parity information is available, the ChipKill decoder proceeds with the same correction procedure as in the baseline flow shown in Figure~\ref{fig: discuss2}(a).
A potential concern is whether this decoupled process offers the same level of protection as the default baseline process. To ensure equivalence, any silent data corruption that bypasses the detection stage in Figure~\ref{fig: discuss2}(b) must also remain undetected or miscorrected in the baseline flow. A detailed analysis of this condition is presented in Section~\ref{sec:evaluation-relia}.

\begin{figure*}[htbp]
  \centering
  \includegraphics[width=0.93\textwidth]{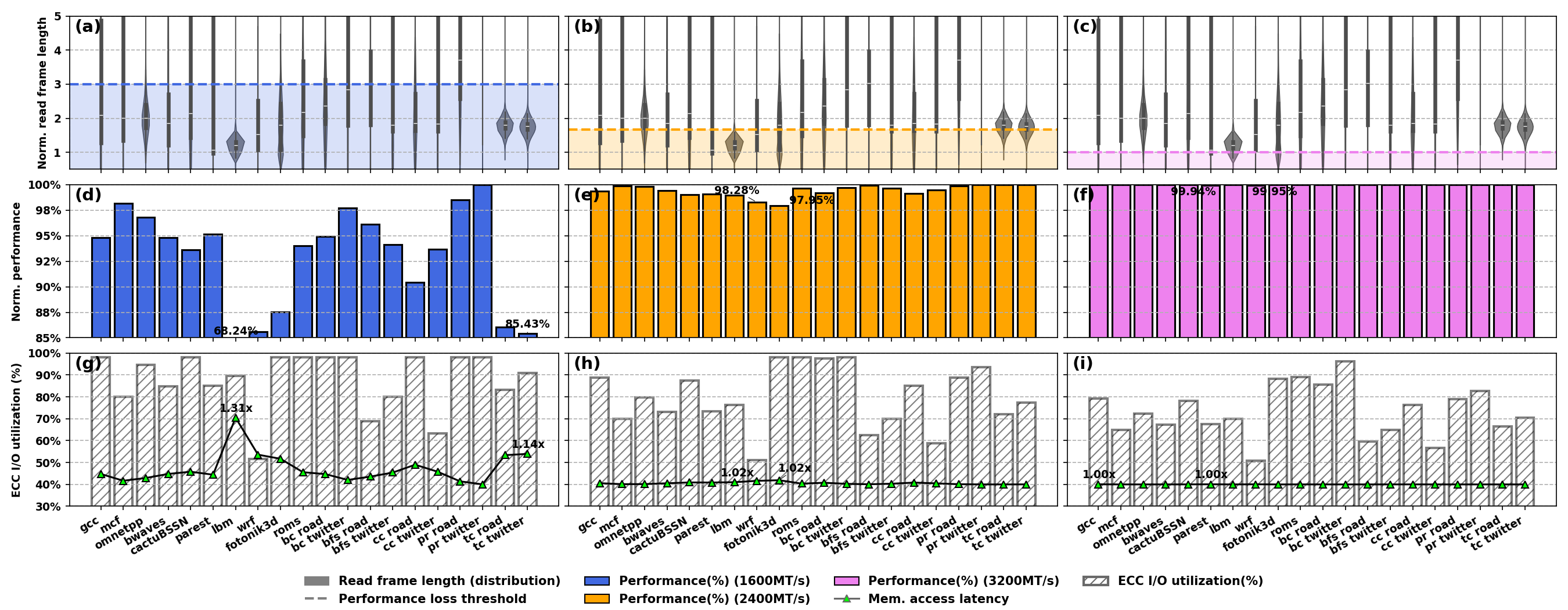}
  \vspace{-0.2in}
  \caption{\textbf{(a)-(c)} Read frame statistics (normalized to write frames) versus varying performance loss thresholds. 
  \textbf{(d)-(f) Overall performance degradation of SCREME-Framewk when employing slow chips with different transmission speeds.}
  \textcolor{black}{(g)-(i) Memory read access latency (lines) and ECC bandwidth utilization (bars).} 
  }
  \vspace{-0.1in}  
  \label{fig: perf1}
\end{figure*}
\section{Evaluation}


\subsection{Methodology}

\subsubsection{Performance}
We evaluate the performance of SCREME using Ramulator 2.0~\cite{luo2023ramulator,kim2015ramulator}, a cycle-accurate DRAM simulator with up-to-date DDR models. Table~\ref{tab:config} lists the main parameters of the simulated system. The memory controller uses open-page and FR-FCFS scheduling policy. The mapping is based on~\cite{wang2020dramdig}, where internal memory hierarchies, e.g., bank groups and banks, are XOR interleaved to minimize conflicts. 
For benchmarks, we use SPEC CPU2017 and GAP. Each multi-programmed SPEC workload is generated using four instances of the same benchmark. We use Pin \cite{luk2005pin} to record virtual address traces for these workloads and page table dumps to translate the recorded traces to physical address traces. For each simulation, we first run two billion instructions to warm up the cache, and then sample one billion instructions using SimPoint~\cite{perelman2003using}. We identify 24 benchmarks, listed in Table~\ref{tab:config}, based on their memory intensity for the evaluation. 
DDR5 timing parameters are obtained from~\cite{micron-ddr5}. DDR5-6400 is used as the baseline module. The default rate of slow DRAM chips is set to half of the baseline, i.e., 3200 MT/s.

\begin{table}[htbp]
\vspace{-0.1in}
\small
\caption{System Configuration and Benchmarks}
\vspace{-0.1in}
\centering
\begin{tabular}{l|ll}
\hline 
 
\hline
            & 32 cores, 4.0GHz, non-inclusive\\
Processor   & L1: 16KB, 64B cacheline, 8-way asso.  \\ 
            & L2: 256KB, 64B cacheline, 8-way asso.  \\ 
            & LLC: 1MB/core, 64B cacheline, 16-way asso. \\ \hline
            
            & Write queue capacity: 32 \\
Memory      & Page management: open-page  \\
Controller  & Scheduler: FR-FCFS \\ 
            & Data buffer latency: 4 \\ \hline
            
            & DDR5-6400, 4GB, $\times$4 I/O width \\
            & 2 channels, 2 ranks, \\
DRAM        & 8 banks-groups, 4 banks/bank-group \\
Memory      & 8Kb local row buffer \\
            & CL-nRCD-nRP-nRAS: 44-44-44-81 \\ 
            & nCCDS-nCCDL-nCCDLWR: 8-16-64 \\ \hline

            & \textbf{SPEC2017 CPU} \\
            & \textbf{int}: gcc, mcf, omnetpp. \textbf{fp}: bwaves, cactuBSSN,  \\
Benchmarks  & parest, lbm, wrf, fotonik3d, roms \\ \cline{2-2}
            & \textbf{GAP}\\
            & bc road, bc twitter, bfs road, bfs twitter, cc road,  \\ 
            & cc twitter, pr road, pr twitter, tc road, tc twitter \\ \hline 
            
\hline
\end{tabular}
\label{tab:config}
\vspace{-0.15in}
\end{table}

\subsubsection{Energy and Cost}

We evaluate DRAM operational power using Micron’s power calculator~\cite{micronpower}, which is based on current (IDD) values measured on actual devices~\cite{micron_ddr5_part_catalog}. 
In addition to operational power, we evaluate the device's marketing prices, which serves as a proxy for representing economic cost. 

\subsubsection{Reliability}
We evaluate reliability using the same metric as~\cite{kim2015bamboo} and~\cite{gong2018duo}. We use Monte Carlo error injection to calculate the error coverage. We refer to~\cite{kim2023unity} to summarize faults into three categories: per-chip Single Bit Fault (SBF), per-chip Double Bit Fault (DBF), and Single Chip Fault (SCF). For SBF and DBF, the chip and bit positions of fault(s) are chosen randomly. For SCF, the chip position and the chip error value are randomly generated.
Based on ECC decoding results, errors are categorized into 3 types: DCE (Detectable and Correctable Errors), DUE (Detectable but Uncorrectable Errors), and SDC (Silent Data Corruption). SDC includes detectable but miscorrected errors and undetectable errors.

\subsubsection{Design Overhead}
Since SCREME is implemented by exploiting under-optimized operations or under-utilized resources, it does not require expensive chip-level modifications. The primary `modification' is to request vendors to provide end users with access to redundant I/O resources.
Additionally, it requires only a minor adjustment to the memory controller. As illustrated in Figure~\ref{fig: discuss1}, this involves augmenting the write buffer with a parity data buffer equivalent to 1/8 of the regular write buffer size. Considering that write buffer sizes typically range from 32 to 128 entries, this adds less than 1KB (with 8B buffered parity per entry) to the controller buffer, which is negligible.


\subsection{Results}

\subsubsection{Basic Performance}
Since SCREME-Framewk inherits and integrates the features of both SCREME-WO and SCREME-I/O, we use SCREME-Framewk to elaborate on the effectiveness of our design and avoid repetition. SCREME-Framework introduces a novel utilization of ECC and I/O resources. Regarding ECC, there are two processes that might impact overall memory performance: (1) the continuous slow writes that are performed by slow ECC chips, and (2) the prolonged error correction process.

We conducted a comprehensive study to unveil the impact of slow writing. As aforementioned, all chips enter write mode synchronously but exit asynchronously. It incurs performance overhead when the extra time required by a burst of slow writes cannot be overlapped by the associated read burst. We refer to the time window occupied by a read burst as the read frame. Figure~\ref{fig: perf1}(a)-(c) show the violin plots, which indicate the distribution of all normalized read frame ($(t_{2}-t_{1})/t_1$ in Figure~\ref{fig: discuss1}). The threshold in the figure represents the additional write frame induced by slow writes ($(t_{1}'-t_{1})/t_1$ in Figure~\ref{fig: discuss1}). The threshold increases as the data rate decreases, bringing read frames closer to the threshold. Performance loss occurs when read frames fall below this threshold. 

Figure~\ref{fig: perf1}(d)-(f) summarize the performance loss. Even under an extreme condition with a 2400 MT/s slow chip (regular chips are 6400 MT/s), the performance and throughput loss is limited to $\sim$3\%. It is not surprising that there is almost no performance overhead ($<$0.1\%) for a 3200MT/s slow chip. 
\textcolor{black}{Figure~\ref{fig: perf1}(g)-(i) present additional metrics, including memory access latency and ECC (I/O) bandwidth utilization. Consistent with the above observations, the latency overhead becomes slightly noticeable ($0.585\%$\ on average) only when \textbf{extremely} slow chips (2400 MT/s) are used. These latency and performance impacts are directly tied to ECC (I/O) bandwidth utilization. Recall that the baseline wastes half of the ECC bandwidth during read accesses (see Section~\ref{sec: write-only}). Our approach repurposes this underutilized bandwidth to accommodate slower chips. However, increased ECC bandwidth utilization also raises the risk of bandwidth contention, which can ultimately degrade performance.}
It is worth noting that all selected benchmarks were filtered by their memory intensity, reinforcing our confidence that a slow chip operating at half the baseline rate will incur nearly zero overhead.


\begin{figure}[htbp]
  \centering
  \vspace{-0.1in}
  \includegraphics[width=0.45\textwidth]{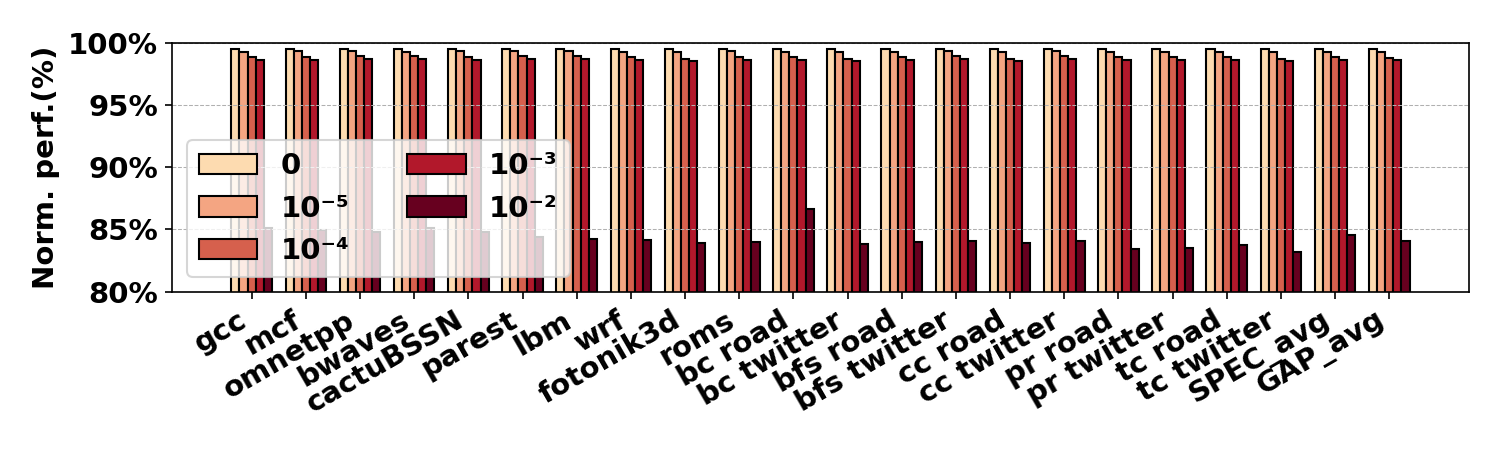}
  \vspace{-0.2in}
  \caption{Performance impact of the decoupled ChipKill process under various error rates.
  }
  \vspace{-0.15in}
  \label{fig: perf2}
\end{figure}

Figure \ref{fig: perf2} shows the performance impact of the prolonged error correction process. Even though SCREME-Framewk needs to perform a dedicated read access to the slow chip with extended latency for each detected error, the performance overhead is negligible due to the relatively small error rate. The overhead only becomes obvious when the error rate approaches $10^{-3}$, a scenario that is practically impossible in real systems, which typically require a Bit Error Rate (BER) below $10^{-6}$~\cite{zhang2018exploring}.

\begin{figure}[htbp]
  \centering
  \vspace{-0.15in}
  \includegraphics[width=0.45\textwidth]{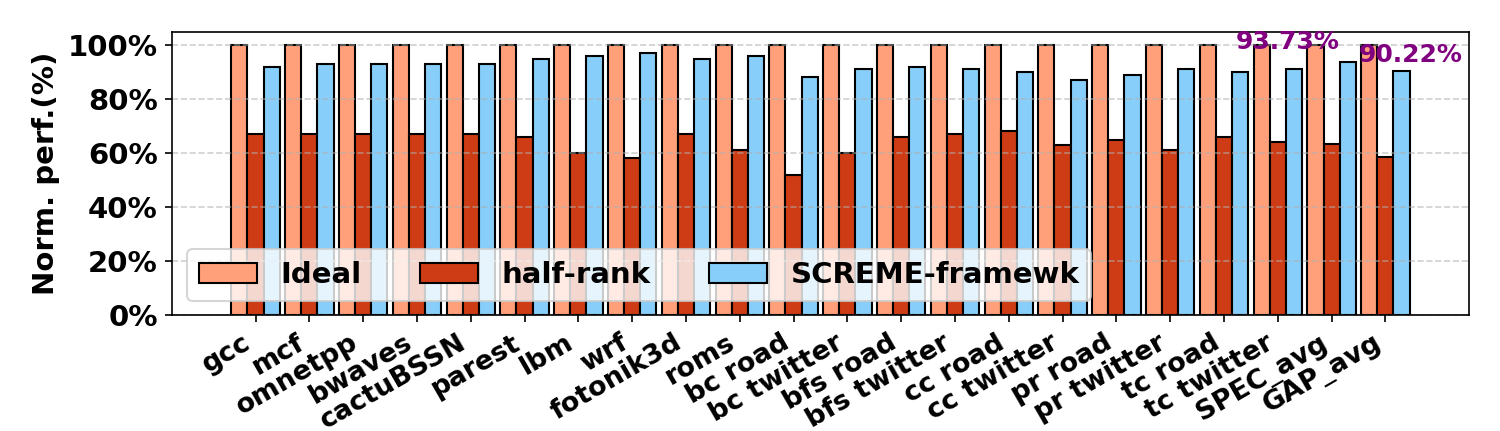}
  \vspace{-0.2in}
  \caption{Performance impact of a module with half-rank failure and SCREME-Framewk.
  }
  \vspace{-0.2in}
  \label{fig: perf3}
\end{figure}

SCREME-Framewk also introduces flexible utilization of I/O resources by reorganizing the rank structure (Figure~\ref{fig: resil_2}). Compared to the simple strategy of using half a rank, our design improves performance by an average of 32\% for SPEC benchmarks and 30\% for GAP benchmarks (Figure~\ref{fig: perf3}. This improvement can be attributed to the efficient reorganization, which fully utilizes the bandwidth without significantly compromising rank-level and bank-level parallelism. The ideal case in Figure~\ref{fig: perf3} assumes all rows in Figure~\ref{fig: resil_2} are available without any failure.

\begin{figure}[htbp]
  \centering
  \vspace{-0.1in}
  \includegraphics[width=0.45\textwidth]{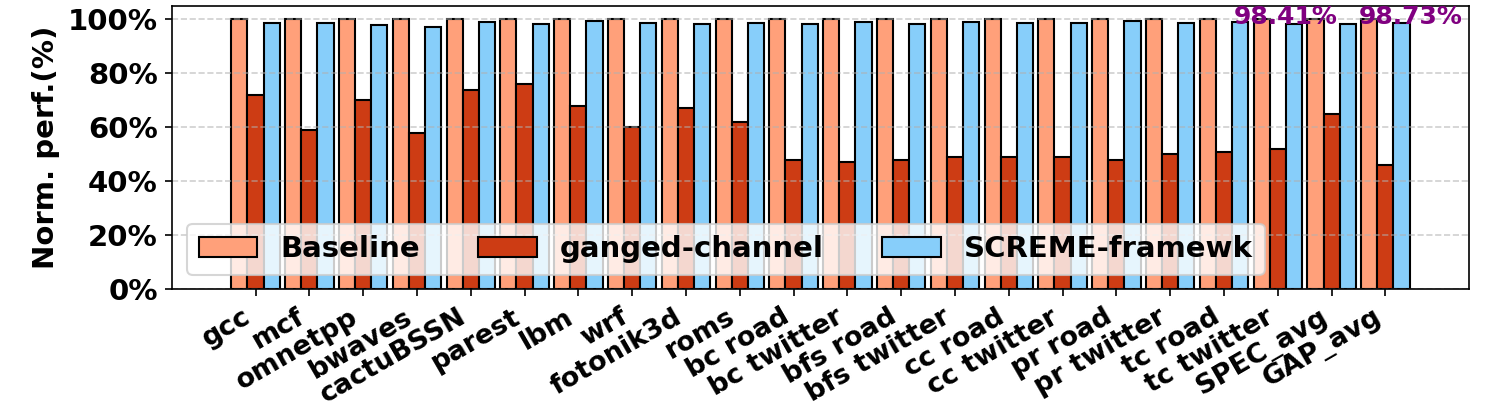}
  \vspace{-0.2in}
  \caption{Performance comparison between the ganged channel approach and SCREME-Framewk.
  }
  \vspace{-0.15in}
  \label{fig: perf4}
\end{figure}

\subsubsection{Comparative Analysis}

We compare SCREME-Framewk with the ganged channel approach used by the DDDC protection scheme, which can achieve the same protection level (Table~\ref{tab:ECC}) as SCREME-Framewk. Figure~\ref{fig: perf4} shows that SCREME-Framewk outperforms the ganged channel approach by 36\% and 53\% for SPEC and GAP, respectively. This highlights the effectiveness of our design. Meanwhile, SCREME-Framewk presents the same performance as the baseline, which has been discussed in Figure~\ref{fig: perf1}.


\begin{figure}[htbp]
  \centering
  \vspace{-0.15in}
  \includegraphics[width=0.45\textwidth]{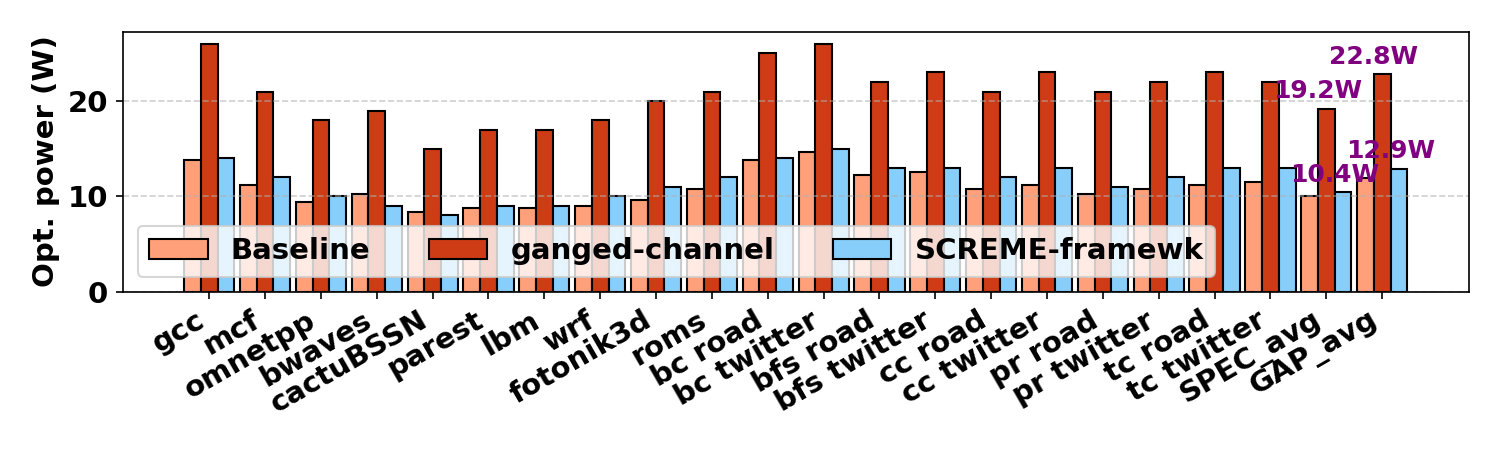}
  \vspace{-0.2in}
  \caption{Operational power comparison between the ganged channel approach and SCREME-Framewk.
  }
  \vspace{-0.1in}
  \label{fig: power1}
\end{figure}

Figure~\ref{fig: power1} presents the operational power consumption of SCREME-Framework. Compared to the baseline, SCREME-Framework incurs a slight power increase due to the involvement of an additional ECC chip. However, this increase is minimal, as the extra ECC chip is configured for write-only operations, thereby avoiding unnecessary read activity.
We also compare SCREME-Framewk to the DDDC approach, which exhibits significantly higher operational power than SCREME-Framework due to its ganged channel scheme, which involves more chips per transfer. Furthermore, the increased access granularity results in the transfer of unnecessary data, exacerbating power consumption.


\subsubsection{\textcolor{black}{Cost Effectiveness}}
\textcolor{black}{To demonstrate the cost-effectiveness of our design, we compare it to a traditional solution that uses regular chips as ECC chips. This traditional approach requires non-trivial interface modifications, as current DDR standards cannot provide additional channel widths for extra ECC chips. In contrast, our design avoids such modifications through the two schemes described in Sections~\ref{sec: write-only} and~\ref{sec: life}. The analysis below assumes an \textit{ideal baseline} in which such an interface extension is available.}

\textcolor{black}{Achieving the same level of reliability with regular chips introduces significant cost challenges, both in terms of purchase and operation.
Figure~\ref{fig: price_info}(a) (derived from Figure~\ref{fig: market}) shows that using slow chips can reduce ECC costs by 2$\sim$6$\times$, depending on the speed gap (1.67$\sim$1.91$\times$). Recall that speed gaps below 2$\times$ lead to negligible performance loss (Figure~\ref{fig: perf1}). 
Given market fluctuations, we also include 2024 pricing data, which confirms the continued cost advantage of slow chips, with 1.4$\sim$3.2$\times$ savings.
Figure~\ref{fig: price_info}(c) shows the operational power consumption. Compared to the \textit{ideal baseline} design that employs regular chips as ECC, our design reduces power by an average of 10$\sim$25\%, depending on the number of additional ECC chips. This reduction results from dedicating the slow chip to write-only operations, which avoids unnecessary reads and reduces power consumption. 
Moreover, this cost-effective feature is critical for scalability, which is increasingly important for next-generation DDR6 (highlighted background in Figure~\ref{fig: price_info}). As aforementioned, ECC chip cost becomes increasingly significant as the number of chips per channel decreases, and our proposed SCREME-Framewk offers a scalable solution to address this challenge.}

\begin{figure}[htbp]
  \centering
  \vspace{-0.1in}
  \includegraphics[width=0.48\textwidth]{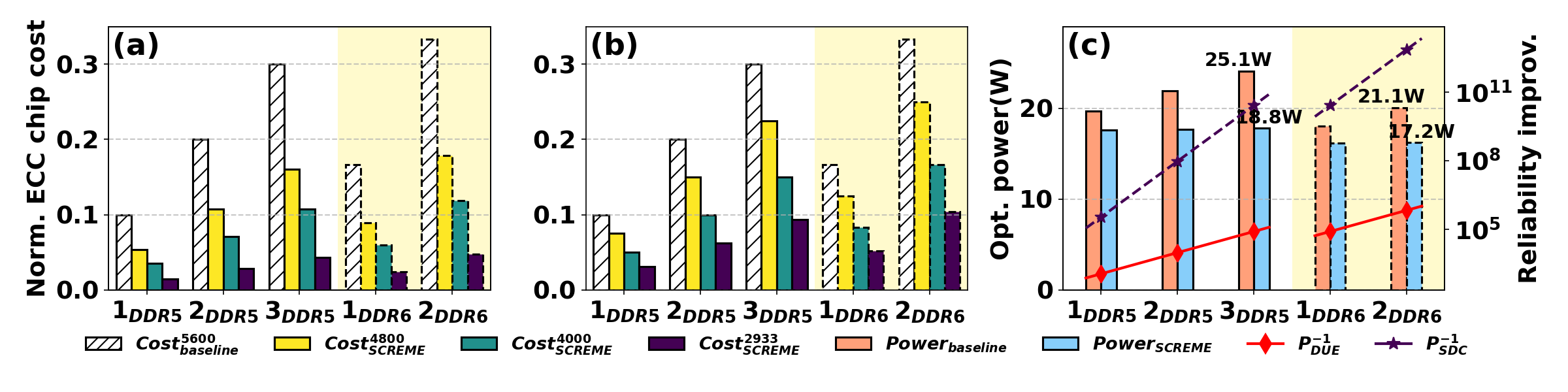}
  \vspace{-0.2in}
  \caption{\textcolor{black}{(a) ECC chip cost comparison using regular chips (DDR-5600) and slower chips (DDR-2933 to DDR-4800) in year 2022.
  (b) ECC chip cost comparison for the same configurations in 2024.
  (c) Operational power consumption comparison using regular chips (DDR-5600) and slower chips (DDR-4000).
  }}
  \vspace{-0.2in}
  \label{fig: price_info}
\end{figure}


\subsubsection{Reliability}
\label{sec:evaluation-relia} 
SCREME-Framewk enhances system resilience by using spare chips to involve more ECC chips or replace failed chips. We assess its reliability along two dimensions: its error-coverage capability and the degree of lifetime extension it affords.

Table~\ref{tab:ECC} presents a reliability comparison of five ECC schemes under ten error scenarios. The first three columns compare the baseline ChipKill with the decoupled ChipKill proposed in this paper. The decoupled ChipKill separates the error detection and correction processes. Based on the results, it is reasonable to conclude that the decoupled ChipKill maintains the same correction capability as the baseline. The slight differences observed are unavoidable due to the stochastic nature of the Monte Carlo process.
To theoretically prove that the decoupled ChipKill is mathematically equivalent to the baseline, it is sufficient to demonstrate that any non-detectable errors in the detection phase remain non-detectable in the correction phase. Given that the ChipKill algorithm (RS code) is polynomial-based. The two check symbols, \{$p_0$, $p_1$\} can be expressed as:
\begin{equation}
    p_0 = a_0 * x_0^0 + a_1 * x_0^1 + ... + a_7 * x_0^7
    \label{eq1}
\end{equation}
\vspace{-0.1in}
\begin{equation}
    p_1 = a_0 * x_1^0 + a_1 * x_1^1 + ... + a_7 * x_1^7
    \label{eq2}
\end{equation}

where \{$a_0$, ..., $a_7$\} represent 8 data symbols. The detect phase uses Equation~\ref{eq1} to detect errors. If multiple errors (say 2 errors) occur in \{$p_0$, $a_0$, ..., $a_7$\}, the equation might still be satisfied. This leads to non-detectable errors (SDC). In such cases, the baseline ChipKill, which uses both Equation~\ref{eq1} and Equation~\ref{eq2}, would attribute the error to $p_1$ based on the minimum Hamming distance~\cite{kim2015bamboo}.
Additionally, the results show that the non-detectable ratio (SDC) in the detection phase is always smaller than that in the correction phase. This is because errors in the detect phase might be miscorrected in the correct phase.

\begin{table}[]
\small
\setlength{\tabcolsep}{0.2mm}{
\caption{A comparison of error detection and correction coverage of different ECC schemes (`-' indicates not applicable)}
\vspace{-0.1in}
\centering
\begin{tabular}{c|c|c|cc|c|c|c}
\hline
\multirow{2}{*}{\begin{tabular}[c]{@{}c@{}}Fault\\ Scenario\end{tabular}} & \multirow{2}{*}{\begin{tabular}[c]{@{}c@{}}Decod. \\ Result\end{tabular}} & \multirow{2}{*}{\begin{tabular}[c]{@{}c@{}}ChipKill\\ only\end{tabular}} & \multicolumn{2}{c|}{Decp. ChipKill} & 
\multirow{2}{*}{\begin{tabular}[c]{@{}c@{}}ChipKill \\ w/ on- \\ die ECC\end{tabular}} &
\multirow{2}{*}{\begin{tabular}[c]{@{}c@{}}Intel\\ DDDC \end{tabular}} & \multirow{2}{*}{\begin{tabular}[c]{@{}c@{}}SCREME \\ Framwk \end{tabular}} \\ \cline{4-5}
 &  &  & \multicolumn{1}{c|}{\begin{tabular}[c]{@{}c@{}}Detect\\ phase\end{tabular}} & \begin{tabular}[c]{@{}c@{}}Correct\\ phase\end{tabular} &  &  &  \\ \hline
SBF(\%) & DCE & 100 & \multicolumn{1}{c|}{-} & 100 & 100 & 100 & 100 \\ \hline
SBF+ & DCE & 87.507 & \multicolumn{1}{c|}{-} & 87.503 & 100 & 100 & 100 \\
SBF(\%) & DUE & 12.101 & \multicolumn{1}{c|}{99.951} & 12.108 & 0 & 0 & 0 \\
 & SDC & 0.392 & \multicolumn{1}{c|}{0.049} & 0.389 & 0 & 0 & 0 \\ \hline
DBF(\%) & DCE & 100 & \multicolumn{1}{c|}{-} & 100 & 100 & 100 & 100 \\
 & DUE & 0 & \multicolumn{1}{c|}{100} & 0 & 0 & 0 & 0 \\
 & SDC & 0 & \multicolumn{1}{c|}{0} & 0 & 0 & 0 & 0 \\ \hline
SCF(\%) & DCE & 100 & \multicolumn{1}{c|}{-} & 100 & 100 & 100 & 100 \\
 & DUE & 0 & \multicolumn{1}{c|}{100} & 0 & 0 & 0 & 0 \\
 & SDC & 0 & \multicolumn{1}{c|}{0} & 0 & 0 & 0 & 0 \\ \hline
SBF+ & DCE & 65.625 & \multicolumn{1}{c|}{-} & 65.626 & 100 & 100 & 100 \\
SBF+ & DUE & 33.303 & \multicolumn{1}{c|}{99.871} & 33.300 & 0 & 0 & 0 \\
SBF(\%) & SDC & 1.072 & \multicolumn{1}{c|}{0.129} & 1.074 & 0 & 0 & 0 \\ \hline
DBF+ & DCE & 46.875 & \multicolumn{1}{c|}{-} & 46.880 & 0 & 100 & 100 \\
DBF(\%) & DUE & 51.458 & \multicolumn{1}{c|}{99.792} & 51.459 & 100 & 0 & 0 \\
 & SDC & 1.667 & \multicolumn{1}{c|}{0.208} & 1.661 & 0 & 0 & 0 \\ \hline
DBF+ & DCE & 8.789 & \multicolumn{1}{c|}{-} & 8.789 & 0 & 100 & 100 \\
DBF+ & DUE & 97.051 & \multicolumn{1}{c|}{99.642} & 97.052 & 100 & 0 & 0 \\
DBF(\%) & SDC & 2.862 & \multicolumn{1}{c|}{0.358} & 2.861 & 0 & 0 & 0 \\ \hline
SBF+ & DCE & 0 & \multicolumn{1}{c|}{-} & 0 & 100 & 100 & 100 \\
SCF(\%) & DUE & 96.863 & \multicolumn{1}{c|}{100} & 96.858 & 0 & 0 & 0 \\
 & SDC & 3.137 & \multicolumn{1}{c|}{0} & 3.141 & 0 & 0 & 0 \\ \hline
DBF+ & DCE & 0 & \multicolumn{1}{c|}{-} & 0 & 0 & 100 & 100 \\
SCF(\%) & DUE & 94.242 & \multicolumn{1}{c|}{100} & 94.236 & 100 & 0 & 0 \\
 & SDC & 5.758 & \multicolumn{1}{c|}{0} & 5.764 & 0 & 0 & 0 \\ \hline
SCF+ & DCE & 0 & \multicolumn{1}{c|}{-} & 0 & 0 & 100 & 100 \\
SCF(\%) & DUE & 77.492 & \multicolumn{1}{c|}{100} & 77.495 & 100 & 0 & 0 \\
 & SDC & 22.508 & \multicolumn{1}{c|}{0} & 22.505 & 0 & 0 & 0 \\ \hline
\end{tabular}
\label{tab:ECC}
 }
\end{table}

It is notable that the error detection and correction coverage of ChipKill diminishes as fault scenarios become more complex. It completely loses its correction capability if errors occur in more than one chip. Additionally, the SDC ratio increases as more errors affect different chips. 
This highlights the need for enhanced protection in memory systems.
As mentioned earlier, modern DDR5 chips employ on-die ECC to enhance overall reliability. The last three columns present a comparison of the baseline ChipKill, DDDC, and SCREME, all of which also utilize on-die ECC. The results indicate that ChipKill benefits significantly from on-die ECC, particularly with a reduction of the SDC ratio to 0. However, in more complex scenarios, such as DBF+SCF and SCF+SCF, the correction capability remains unchanged. More importantly, on-die ECC could turn some originally correctable errors into uncorrectable ones, such as the DBF+DBF scenario. This occurs because on-die ECC, due to its limited correction capability, may miscorrect errors, leading to an amplification of errors.
On the contrary, DDDC and our proposed SCREME-Framewk can successfully handle two sequential chip corruptions by opportunistically retiring a failing chip~\cite{kim2015bamboo}. Given that SCREME-Framework maintains good scalability for parity storage, we believe that with greater availability of slow chips in the near future, it can be augmented with more spare space, thereby achieving a stronger capability to tolerate more errors.

\begin{figure}[htbp]
  \centering
  \vspace{-0.13in}
  \includegraphics[width=0.48\textwidth]{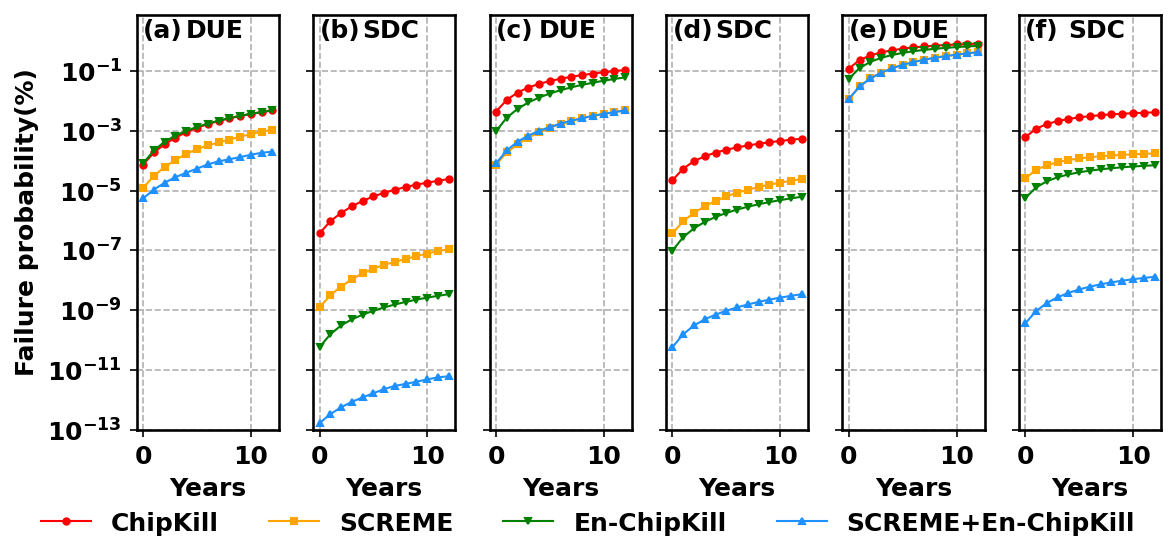}
  \vspace{-0.3in}
  \caption{DUE (a) and SDC (b) failure probabilities of a DDR5 module with various protection schemes. DUE (c) and SDC (d) failure probabilities with one failed chip in the module. \textcolor{black}{DUE (e) and SDC (f) failure probabilities with one failed chip in the module and 10x higher FIT.}
  }
  \vspace{-0.15in}
  \label{fig: lifetime}
\end{figure}

In the lifetime evaluation, we introduce an enhanced ChipKill scheme that doubles the codeword size~\cite{bamboo}. As depicted in Figure \ref{fig: lifetime}(a), SCREME-Framewk achieves a lower DUE probability than the baseline and enhanced ChipKill schemes. In terms of SDC, it falls between the baseline and enhanced ChipKill (Figure~\ref{fig: lifetime}(b)). More importantly, SCREME-Framewk can be combined with the enhanced ChipKill to deliver consistently superior protection, since the two approaches are orthogonal.
When a chip in a module permanently fails, both baseline and enhanced ChipKill protections deteriorate rapidly (Figure~\ref{fig: lifetime} (c) and (d)). In contrast, SCREME-Framewk continues to provide relatively strong protection.
\textcolor{black}{As the error rate continues to rise (Figure~\ref{fig: lifetime}(e) and (f)), such as in space or high-radiation environments, none of the evaluated protections remains reliable. This underscores the need for additional ECC chips under extreme conditions and highlights SCREME-Framewk as a low-cost, scalable solution.}

\section{Conclusion}


In summary, we propose three designs to enhance the resilience of memory subsystems. The first design leverages the suboptimal ECC process to accommodate cost-effective, but low-performance memory chips. The second design exploits underutilized I/O resources to tolerate connection failures in memory modules. The third design combines these approaches into a practical framework for large-scale server-class memory modules. 



\bibliographystyle{ACM-Reference-Format}
\bibliography{bibs/proposal,bibs/xin,bibs/weidong,bibs/yanan}

\end{document}